\documentclass[article]{revtex4-1} 
\usepackage[T1]{fontenc}
\usepackage[utf8]{inputenc}
\usepackage[english]{babel}
\usepackage{amsmath,amssymb,amsfonts}
\usepackage{graphicx,graphics}
\usepackage[caption=false]{subfig}
\usepackage{multirow}
 \usepackage{dcolumn}
\usepackage{bm}
\usepackage{color}
\usepackage{amsthm}             
\usepackage{amssymb}
\usepackage{mathrsfs}
\usepackage{dcolumn}
\usepackage{bm}
\usepackage{epsfig}
\usepackage{epstopdf}
\usepackage{float}
\usepackage{multirow}

\begin{document}
\title{Approximate symmetries of long-range Rydberg molecules including spin effects}
\author{H. Rivera-Rodr\'{\i}guez}\email{hrivera@ciencias.unam.mx}
\affiliation{Instituto de F\'{\i}sica, Universidad Nacional Aut\'onoma de M\'exico, Apdo. Postal 20-364, 01000 Cd. de M\'exico, M\'exico}

\author{R. J\'auregui}\email{rocio@fisica.unam.mx}
\affiliation{Instituto de F\'{\i}sica, Universidad Nacional Aut\'onoma de M\'exico, Apdo. Postal 20-364, 01000 Cd. de M\'exico, M\'exico}

\begin{abstract}
An operator that generates an approximate symmetry of long-range Rydberg molecules (LRRMs) formed by two alkali atoms, one in a Rydberg state and the other in the ground state, is identified. This is first done by evaluating the natural orbitals associated to a variational calculation of the binding wave function within the Born-Oppenheimer description of the molecule including  $s-$ and $p-$ Fermi pseudopotential and the hyperfine structure energy terms. The resulting orbitals with highest occupation number are shown to be identical to those obtained by a perturbative model for  high angular momentum --trilobite and butterfly-- LRRMs. Whenever the slight dependence of the quantum defects of the Rydberg electron on its total momentum $\vec j = \vec \ell +\vec s_1$ can be neglected, the symmetry operator of the
high angular momentum LRRMs orbitals is identified as the sum of the spin of the Rydberg electron $\vec s_1$, spin of the valence electron $\vec s_2$  and
the spin of nucleus $\vec i$  of the ground state atom, $\vec {N} =\vec {s_1} + \vec {s_2} + \vec {i}$. The spin-orbitals that diagonize $\vec{N}$ define  compact basis sets for the description of LRRMs beyond the aforementioned approximations. The matrix elements of the Hamitonian in these basis sets have simple expressions, so that  the relevance of triplet and singlet contributions can be directly estimated.
The expected consequences of this approximate spin-symmetry on the spectra of LRRMs are briefly described. 
\end{abstract}

\maketitle
\section{Introduction}
A highly excited electron of a Rydberg atom scattered by one or more ground-state atoms can lead to the formation of long-range Rydberg molecules (LRRMs). Novel structure properties of ultracold LRRMs are their large dipole moments present even for homonuclear diatomic realizations~\cite{Greene2000, Booth2015, Rivera2021}, relatively long lifetimes~\cite{Niederprum2016}, huge bond lengths with a vibrational dynamics in the microsecond timescale~\cite{YiQuanZou2023}, and binding energies on the meV scale. Experimental generation of these molecules~\cite{Bendkowsky2009,Bendkowsky2010, Bellos2013, Krupp2014, Booth2015,Sassmannhausen2015,DeSalvo2015, Niederprum2016,Schlagmuller2016,Niederprum2016nature, Kleinbach2017, Camargo2018, MacLennan2019, Whalen2019, Engel2019, Whalen2019_2, Ding2019, Deiss2020, Whalen2020, Peper2020,  Peper2021, Peper2023} has confirmed their predicted  high sensitivity to the scattering phase shifts which determine the Fermi pseudopotential~\cite{Fermi1934, Omont1977} that describes the primary chemical bonding mechanism~\cite{Greene2000,Khuskivadze2002,Hamilton2002,Greene2023}.  Their interaction with external electric \cite{Kurz2013, Kurz2014, Eiles2017_b, Hummel2021electric},  magnetic \cite{Kurz2014, Hummel2018, Hummel2019} and electromagnetic waves  \cite{Hollerith2022} offers an ideal scenario for the study of  interesting dynamics such as induced remote spin flips~\cite{Niederprum2016}, beyond Born-Oppenheimer approximation molecular physics~\cite{Hummel2021, Hummel2023,Srikumar2023,Schlagmuller2016}, and scattering of negative ions at low temperatures ~\cite{Engel2019}.

For alkali atoms and diatomic LRRMs, it has been recognized that the  orbital angular momentum ($\vec \ell$) of the scattered Rydberg electron and the intrinsic electronic  angular momenta of both  the scattered Rydberg and the ground-state atom, $\vec s_1$ and $\vec s_2$ respectively, play an essential role not only in the resulting phase shifts but also in the fine and hyperfine structure of the system~\cite{Anderson2014,Eiles2017, Greene2023}. Thus, the quantum numbers associated to the  preparation of the atomic Rydberg state previous to the scattering process define control parameters of  the properties of the generated LRRM including its spectroscopy. That is, LRRMs provide an unique platform to perform precise scattering experiments in the low-energy regime with clearly identified actuating variables.
 
Here we are mainly interested in understanding the general features of the spectroscopy of high-$\ell$ diatomic Rydberg molecules. This requires the identification of simple electronic wave functions that incorporate spin effects.  Due to simultaneous hyperfine and Rydberg electron--ground state atom interaction, the full Hilbert space can not be separated {\it a priori} into subspaces of well-defined hyperfine and singlet/triplet scattering states.  A good quantum number for the system is the total angular momentum projection onto the internuclear axis $\Omega=m_j+m_2+m_i$; it involves the projections of the total angular momentum of the Rydberg electron, $m_j=m_{\ell}+m_1$, and those of the electron spin $m_2$ and the nuclear spin $m_i$ of the ground state atom. Another quantum number is provided by the principal quantum number $n$ of the  Rydberg electron for an asymptotic separation between the ground and Rydberg atoms. 
The spin interaction $\vec s_1\cdot \vec s_2$ between the electrons is directly manifested in the scattering phase shifts and the corresponding Fermi pseudopotential. Simultaneously the magnetic interaction between the electron $\vec{s_2}$ and nuclear $\vec{i}$ spins of the ground state atom, $\vec s_2\cdot \vec i$ is evident in the hyperfine structure of the system. These features prevent the identification of an operator that exactly generates a spin-symmetry of the LRRM.

Nevertheless, the identification of an operator $\vec{N}$ that {\it approximately} generates such a symmetry can be used to distinguish interesting phenomena to be expected both in the spectroscopy realm and in the spin-dynamics context. We employ several techniques to such an end; it results that all of them point to the same final result. We start making a variational treatment of the LRRM within the Born-Oppenheimer approximation. The basic spin-orbitals are defined by the exact symmetry operator $\hat{\Omega}$. From such a calculation a set of natural orbitals with the highest occupation numbers is determined. We show that the spin structure  of the LRRM can always be written in terms of few simple configurations. Simultaneously, a perturbative treatment of both Fermi pseudopotential and hyperfine Hamiltonian is done. The resulting spin-orbitals are shown to coincide with those obtained as natural orbitals generated by the variational treatment. These orbitals are eigenfunctions of the operator $\vec{N}=\vec s_1 + \vec s_2 + \vec i$ which is then identified as a generator of an approximate symmetry of the system. Finally we discuss the consequences on LRRMs behavior.

Most calculations are reported for $^{39}$K molecules, though they were performed also for $^{87}$Rb. For both cases, the variational results are similar to those that have been reported in the literature~\cite{Eiles2018, Niederprum2016,Peper2020}. A comparison of the role of the operator $\vec{N}$ for   $^{39}$K and $^{87}$Rb in the description of LRRMs is made.

In the next Section,  the Hamiltonian that models the system and the expressions of the matrix elements of its terms in the uncoupled basis is revisited. Then, the three approaches to study the approximate potential energy curves (PECs) and wave functions associated to the electronic structure are described. The results of the implementation of those approaches for   high-$\ell$ LRRMs of  $^{39}$K and  $^{87}$Rb are reported and briefly discussed. In Section IV, the symmetries of the spin-orbitals that yield a compact representation of the electronic wave function according to the results of the previous Section are identified. The matrix elements of the Hamiltonian in this basis set are worked out in Section V. Finally, in Section VI, the consequences of the approximate symmetry of the electronic wave functions on the spectroscopy of LRRMs are discussed, and the conclusions are given.
 
\section{Theoretical framework}\label{sec:theo}
Consider LRRM constructed from a Rydberg atom coupled to a ground state alkali atom. Within the Born-Oppenheimer approximation and the pseudopotential scheme, the Hamiltonian that describes the dynamics of the highly excited electron located at $\vec{r}$ relative to the Rydberg core for a given separation $R$ of the atomic nuclei including  relativistic effects is
\begin{equation}
\hat{H}(\vec{r}; R)= \hat{H}_{\mathrm{Ryd}}+\hat{H}_{\mathrm{pol}} +\hat{V}_{\mathrm{Fermi}}+\hat{H}_{\mathrm{HF}}.
\label{eq:hspin}
\end{equation}
Here $\hat{H}_{\mathrm{Ryd}}$ is the  Rydberg  Hamiltonian that takes into account the attraction of the electron to the atomic core and its spin-orbit interaction, whose effect can be encompased through a dependence on the angular momentum $\vec j=\vec \ell + \vec s_1$ of the quantum defects. The hyperfine structure of the Rydberg atom is not considered, as this contribution is much smaller than the rest of the interactions for all alkali atoms studied. The term $\hat{H}_{\mathrm{pol}}= -{\alpha_p}/{2R^4}$ corresponds to the polarization potential between the Rydberg core and the scattered atom. The Fermi pseudopotential $\hat{V}_{\mathrm{Fermi}}$ incorporates the electron scattering channels up to the $p$-wave; their strengths are  parametrized by the relative wave number $k=\sqrt{2/R -1/n_H^2}$, the total $\vec S = \vec s_1 +\vec s_2$ electronic spin (triplet or singlet), the orbital angular momentum of the scattered electron with respect to the ground state atom $\vec L$, and the total angular momentum $\vec J =\vec L +\vec S$. 
In this work, the values of the corresponding scattering lengths and volumes for each $^{2S+1}L_J$ configuration are evaluated using  the phase shifts reported in \cite{Eiles2018,Khuskivadze2002}. The ground state atom hyperfine term is  $\hat{H}_{\mathrm{HF}}=A_{\mathrm{HF}} \, \vec{i} \cdot \vec{s}_2$, where as before $\vec{i}$ is the nuclear spin operator of the ground state atom. The constant $A_{\mathrm{HF}}$ determines the intensity of the hyperfine interaction for each element. The $\alpha_p$ and $A_{\mathrm{HF}}$ parameters used in our calculations correspond to those reported in \cite{Eiles2017,Eiles2018}.  

We choose to represent the Hamiltonian in a basis formed by the eigenstates of the Rydberg atom $|n\ell jm_j\rangle$ and the uncoupled nuclear and electronic spin states of the ground state atom $| s_2 m_2;i m_i\rangle$ where $m_2$ and $m_i$ are the projections of the electronic $s_2$ and nuclear $i$ spin respectively. Therefore, we seek to find the representation of the Hamiltonian in the basis $| n  \ell j m_j \rangle \otimes | s_2 m_2;i m_i\rangle$. 

Once the ground state atom species has been determined, the value of $i$ is fixed. This allows us to replace $| s_2 m_2;i m_i\rangle$ with $|m_2, m_i\rangle$ to simplify the notation without risk of confusion. The nuclear spin is $i=3/2$ for $^{87}\mathrm{Rb}$ and $^{39}\mathrm{K}$. The matrix elements of $\hat{H}_{\mathrm{Ryd}}$,  $\hat{H}_{\mathrm{pol}}$ and $\hat{H}_{\mathrm{HF}}$ in the uncoupled basis are straightforward to obtain. 

The model developed in \cite{Eiles2017} allows to write the Fermi pseudopotential in a way that correctly incorporates the dependence on the total electronic spin $\vec{S}$ and angular momentum $\vec{L}$. The Fermi pseudopotential is diagonal in the basis formed by states of relative angular momentum to the perturber $|LSJ M_J \rangle\equiv\vert \beta\rangle$. The $\beta$ quantum numbers are incompatible with the $\eta=\lbrace n,\ell,j, m_j \rbrace$ quantum numbers characterizing the eigenstates of the Rydberg electron. To find the matrix elements of the Hamiltonian operator of Eq.~\eqref{eq:hspin} in the $| n  \ell j m_j \rangle \otimes |m_2, m_i\rangle$ basis, it is necessary to perform an expansion of the electronic wave function about the position of the ground state. A frame transformation  matrix $\mathbb{\mathcal{A}}$ 
changes the coordinates and quantum numbers between the Rydberg core and the perturber atom. Written explicitly,
\begin{eqnarray}
\mathcal{A}_{\eta s_2m_2,\beta}&=& \sum_{M_L=-L}^{L} \sqrt{\frac{4 \pi}{2\ell +1}} C_{\ell M_L,s_1m_j-M_L}^{jm_j} Q_{LM_L}^{n\ell j}(R)  C_{s_1m_j-M_L,s_2 m_2}^{S m_j-M_L+m_2} C_{L M_L,S m_j-M_L+m_2}^{J m_j+m_2},
\label{eq:AA}\\
Q_{0 \,0}^{n\ell j}(R)&=& \sqrt{\frac{2\ell +1}{4 \pi}} \frac{f_{n\ell j}(R)}{R},\label{eq:qespin00}
\\
Q_{1 \,0}^{n\ell j}(R)&=& \sqrt{\frac{2\ell +1}{4 \pi}} \left. \frac{d }{dr} \left( \frac{f_{n\ell j}(r)}{r} \right) \right|_{r=R} \label{eq:qespin10}\\
Q_{1 \, \pm 1}^{n\ell j}(R)&=& \sqrt{\frac{(2\ell+1)(\ell+1) \ell}{8 \pi}} \frac{f_{n\ell j}(R)}{R^2} \label{eq:qespin11}
\end{eqnarray}
where $f_{n\ell j}(R)$ are the radial eigenstates of the Rydberg atom.

The scattering matrix $\mathbb{V}_{\mathrm{Fermi}}$ is diagonal in the nuclear angular momentum $m_i$ and can be written as
\begin{eqnarray}
\mathbb{V}_{\mathrm{Fermi}}&=& \mathbb{ \mathcal{A}} \times \mathbb{U} \times \mathbb{\mathcal{A}^{\dagger}}.
\label{eq:Vfinal}\\
U_{\beta, \beta^\prime }&=& \delta_{\beta \beta^\prime} \frac{(2 L +1)^2}{2} a(L \, S \, J,k),
\end{eqnarray}
where $a(LSJ,k)$ is the scattering length (volume) corresponding to the scattering channel with quantum numbers $L, S$, and $J$. It depends implicitly on the nuclear separation $R$ through the wave number $k(R)$.\\

Note that the rotational symmetry along the internuclear vector guarantees that a  good quantum number for the system is the total angular momentum projection $\Omega=m_j+m_2+m_i$. As a result, the matrix of $\hat{H}$ is block diagonal in $\Omega$, so it is possible to solve for each value of $\Omega$ independently. 

\section{Adiabatic description of LRRMs}
We are interested in obtaining compact representations of the electronic wave functions of LRRMs that allow the understanding of the spin structure within the Born-Oppenheimer approximation. We consider three different approaches to obtain the approximate potential energy curves (PECs) and wave functions  for $^{39}$K and $^{87}$Rb homonuclear LRRMs:   
\begin{itemize}

\item[(i)]The diagonalization of the Hamiltonian matrix in a given truncated basis for each value of the internuclear distance $R$. Within this basis, the spin quantum numbers are finite and their incorporation is direct, while the principal quantum number of the hydrogen-like orbitals is used to define a truncated basis. The PECs are base dependent; it has been found that there is not a well defined convergence in this diagonalization method \cite{Greene2023, Fey2015}.  
Nevertheless, it has also been recognized that using two $n$-manifolds below and one above the level of interest usually produces the best agreement with other methods \cite{Eiles2017} and well as with experimental spectroscopic results \cite{Booth2015, Niederprum2016nature, Kleinbach2017}.
Having this in mind, our calculations use a basis that includes the four hydrogenic manifolds $\lbrace n_H-2, n_H-1, n_H,n_H+1 \rbrace$ and all Rydberg states whose energy lies between $E_{n_H-2}$ and $E_{n_H+1}$.  This approach yields an accurate representation of the electronic wave functions that nevertheless is, in general, not compact.

\item[(ii)] Using the eigenstates obtained in (i), a density matrix is constructed for each $R$ value. The corresponding natural orbitals \cite{Lowdin1956} (eigenstates) and occupation numbers (eigenvalues) of such a reduced density matrix are obtained. The natural orbitals with highest occupation numbers are taken as an approximation to the electronic wavefunction. In all the cases we have considered, two natural orbitals are enough for achieving an occupancy number around 0.99.

\item[(iii)] A perturbative approach taking as reference $\hat{H}_0 =\hat{H}_{\mathrm{Ryd}}+\hat{H}_{\mathrm{pol}}$
is performed. Since the eigenstates of $\hat{H}_0$  for each principal quantum number $n_0$ are nearly degenerate, a new set of orbitals is obtained by diagonalizing $\hat{H}-\hat{H}_0$. 
\end{itemize}

We compare the results of (ii) and (iii). Their accuracy is estimated from their fidelity with the
exact eigenfunctions obtained by (i), and their comparison with the respective numerical PECs. We have found that (ii) and (iii) give numerically equivalent results, with a high fidelity to the exact electronic function. Even more important, for the high-$\ell$ molecular states the  relevant (ii) and (iii) spin-orbitals are shown to be eigenstates of an operator
\begin{equation}
\vec {N} =\vec s_1 + \vec s_2 + \vec i.
\end{equation} 
 
As a consequence, $\vec N$ defines the generator of an approximate symmetry of the system. The degree of approximation of this symmetry depends on the scattering parameters of the ground state atom. Particularly the splitting between $^{3}P_J$ phase shifts. For this reason, the symmetry is always present in molecular states resulting from $s-$wave scattering but its range of validity on $p-$wave scattering states depends on the atomic species of the ground state atom.  

\subsection{Numerical diagonalization}
Here we present the results obtained through numerical diagonalization using the full Hamiltonian given by Eq.~\eqref{eq:hspin} including all scattering channels ($s-$, $p-$wave interactions, singlet and triplet electronic configurations). The Fermi-Omont effective zero-range interaction terms
in  Eq.~\eqref{eq:hspin} depend on the scattering phase shifts. Here we consider the values reported in Ref.~\cite{Eiles2018} for $^{39}$K and those in Ref.~\cite{Khuskivadze2002} for $^{87}$Rb. The phase shifts for $^{39}$K reported in \cite{Eiles2018} are non relativistic and therefore are not $J-$ dependent. Since the $^3P_J$ splitting in light alkali-metal atoms scales as $Z^4/n^3$ \cite{Bahrim2001}, for K this fine structure splitting is estimated to be 182 $\mu$eV \cite{Peper2021} and therefore neglecting the spin-orbit coupling in the $e-$K scattering is reasonable and sufficiently valid \cite{Eiles2018, Peper2021}. Based on this, we take all three $^3P_J$ scattering phase shifts to be equal for our calculations in $^{39}$K.

Since only the first two partial waves ($L \leqslant 1$) are considered the states that will be modified are those with $|m_j| \leqslant 3/2$. With the possible spin quantum number values of  $^{39}$K and $^{87}$Rb  ($i=3/2$), we have the cases $|\Omega|= \frac{1}{2},\frac{3}{2},\frac{5}{2},\frac{7}{2}$ . For states around $n_H=35$, the number of elements in the basis of each block varies from 2340 in the case of $|\Omega|=1/2$ to 330 for maximum $|\Omega|$.

It must be mentioned that interesting results derived from an alternative fully spin-dependent Green’s function method have been recently reported~\cite{Greene2023}.
In the formalism used in that work the phases of the wavefunction are directly manipulated, avoiding a diagonalization procedure.

As it is customary in this type of Rydberg-ground state molecules,  two different classes of molecular states are identified. The first type is the low angular momentum molecule in which the Rydberg electron has a well-defined $\ell \leqslant 2$ orbital angular momentum. The second type is the high$-\ell$  molecules. In this class of molecule the electronic state is formed by a mixing of high$-\ell$ hydrogenic states. Within this high$-\ell$ class there are two different subfamilies of states. The trilobite and butterfly states originated by $s-$wave and $p-$wave scattering respectively. Here we focus on high orbital angular momentum molecular states. The minimun orbital angular momentum value $\ell_{\mathrm{min}}$ from which we consider that we speak of high$-\ell$ states is determined by the quantum defects. For the atomic species studied in this work we use $\ell_{\mathrm{min}}=3$.

\begin{figure}[ht!]
	\centering	
	\includegraphics[width=0.93\columnwidth]{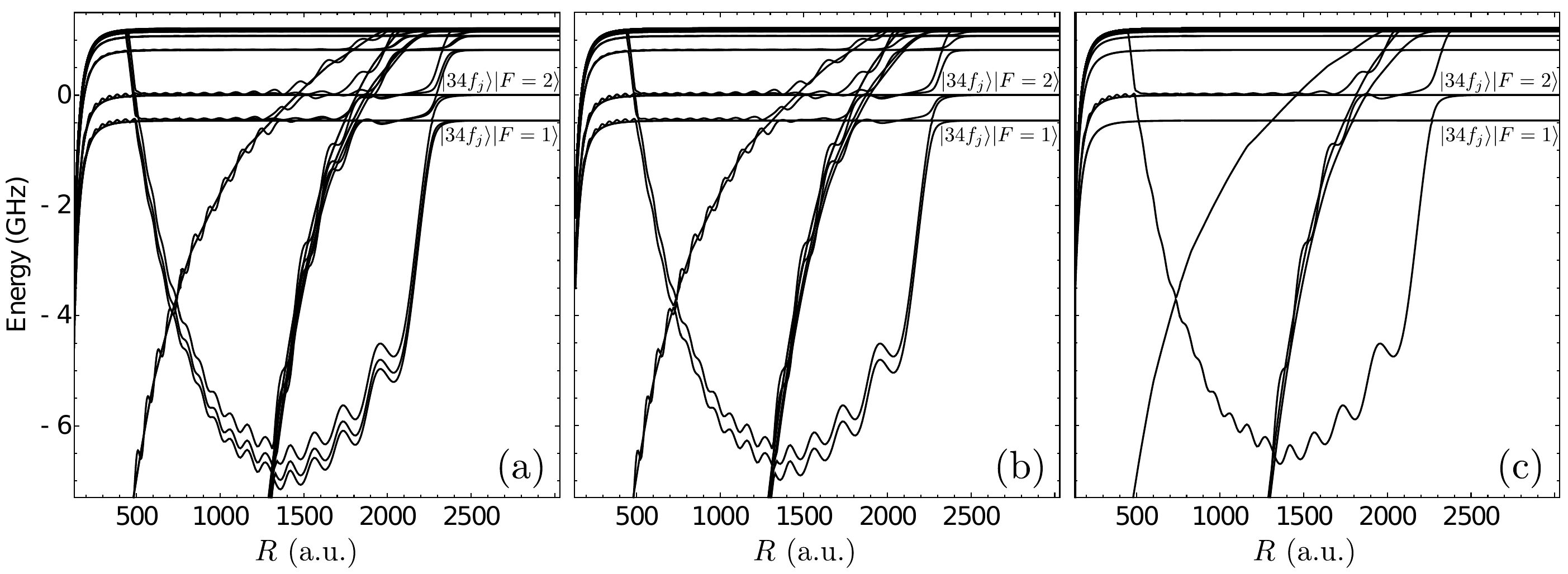}
	\caption{PECs for $n=34$ triplet trilobite $^{39}$K LRRM as a function of the internuclear distance $R$ for (a) $\Omega=1/2$, (b) $\Omega=3/2$ and (c) $\Omega=5/2$. The energy scale is relative to $|34 f_{5/2} \rangle |2 \rangle$.}	
	\label{fig:Ktrilospin}
\end{figure}

Figure~\ref{fig:Ktrilospin} shows a set of numerical PECs for triplet dominated trilobites in the vicinity of states $|34 (\ell \geqslant \ell_{\mathrm{min}}) \rangle  |F \rangle$ for the possible values of $\Omega$ in $^{39}$K$_2$ LRRMs. The hyperfine structure adds multiplicity that increases as the value of $|\Omega|$ decreases. In these trilobite PECs, the effect of the shape resonance is evident, creating a set of avoided crossings and sharp drops around $R=800 \, a_0$ and $R=1300 \, a_0$ where $a_0$ is the Bohr radius. For the extreme value $\Omega=7/2$ there are no PECs that support bound states. The depth of the PECs is on the order of GHz and, despite the effect of the shape resonance, trilobite PECs are formed in a dominant way by $s$-wave scattering.  The shape resonances modify them through the appearance of avoided crossings. Another notable difference compared to low angular momentum PECs \cite{Peper2020, Niederprum2016} is that all trilobite curves originate almost entirely from one spin-character scattering length ($S=0$ or $S=1$). The triplet dominated PECs correspond to bound states. The splitting into multiple PECs occurs mostly as a result of the hyperfine interaction of the ground state atom. For different values of $\Omega$, both trilobite PECs with well-defined $F$ and PECs that exhibit a combination of hyperfine states are present.

\begin{figure}[ht!]	
	\centering	
	\includegraphics[width=0.93\columnwidth]{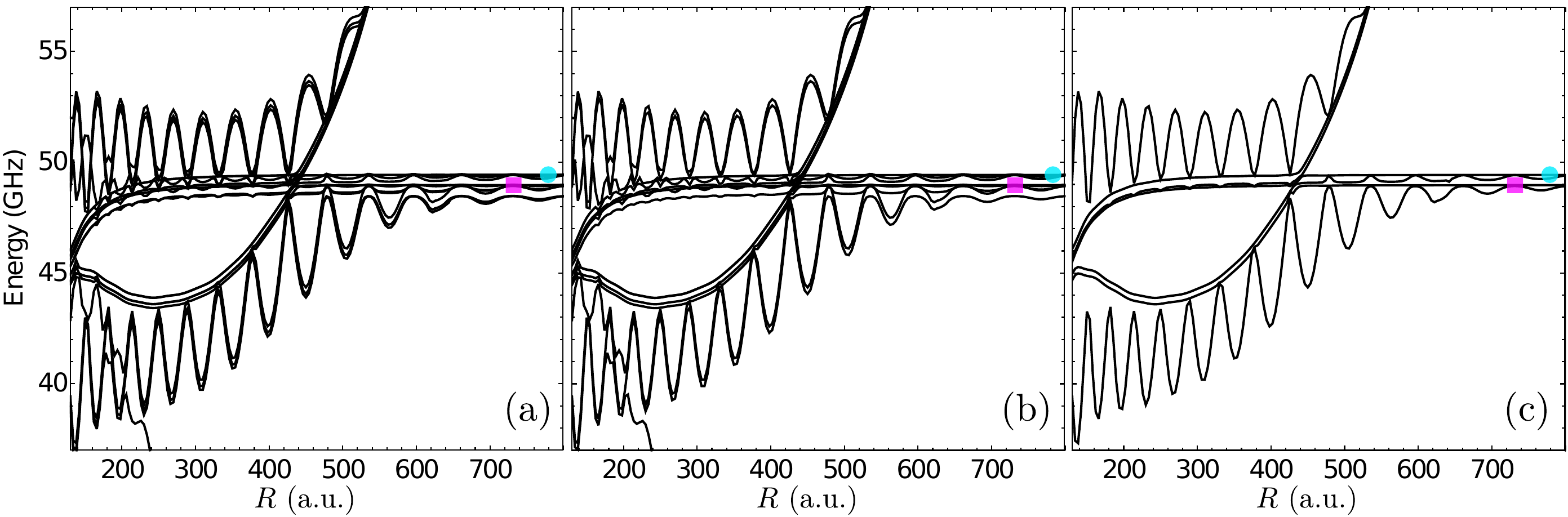}
	\caption{PECs for $n=34$ dominated triplet butterfly $^{39}$K LRRM as a function of the internuclear distance $R$ for (a) $\Omega=1/2$, (b) $\Omega=3/2$ and (c) $\Omega=5/2$. The energy scale is relative to the $|34 f_{5/2} \rangle |2 \rangle$ state. The blue circle corresponds to the asymptotic energy of the $\vert 36 \, p_{3/2}\rangle\vert 2\rangle$ state and the pink square to that of $\vert 36 \, p_{3/2}\rangle \vert 1\rangle$.}	
	\label{fig:Kmariposaspin}
\end{figure}

Some butterfly PECs are shown in Fig.~\ref{fig:Kmariposaspin}. Butterfly molecules with principal quantum number $n$ are bonded in the vicinity of $(n+2)p$ states as a consequence of the quantum defect value ($\mu_1 \approx 1.7$) for Potassium. We observe several deeper wells (with depths of GHz) that are regularly spaced and support the existence of multiple vibrational levels for nuclear separations $R\approx 100 \,a_0-400 \,a_0$. In the same way as in the trilobite case, for smaller values of $|\Omega|$ we have more multiplicity of butterfly PECs. The PECs presented here are consistent with the results reported in Ref.~\cite{Eiles2018}.

In the case of $^{87}$Rb, there is a observable splitting of the $^{3}P_J$ phase shifts. This introduces richness and additional structure to the PECs in this atomic species, as can be seen in the insets of Fig. \ref{fig:NRb}. Later, we will analyze the origin of said additional structure in terms of the symmetries of the system.

\subsection{Numerical Natural Orbitals}
For high$-\ell$ trilobite states, we find that the natural orbitals with the highest occupancy numbers are built from spin-orbitals with a single principal quantum number $n$  ($n=34$ in the particular example corresponding to $^{39}$K) and with all possible values of $\ell \geqslant \ell_{\mathrm{min}}$ and $j$ with projection $m_j=\pm 1/2$. Additionally, for all cases, the electronic part can be separated into the same two $R-$dependent states $|u_{\pm} ^{(n)}(R) \rangle$ defined by the projection $m_j$. For example, for $\Omega=3/2$ and triplet dominated trilobites, in the spatial region where we expect bound states, each of the two states can be written as
\begin{equation}
|\psi_{X}^T\rangle \approx \sqrt{\lambda_{X+}^T(R)} |u_{+}^{(n)} (R)\rangle (A_X \, |1 \, 1 \rangle+B_X \, |2 \, 1 \rangle)+ \sqrt{\lambda_{X-}^T(R) } |u_{-}^{(n)}(R) \rangle  \, |2 \, 2 \rangle,
\label{eq:trilo_nat1}
\end{equation}
with $X=1,2$ labeling each of the states. The values $A_X$ and $B_X$ that multiply the hyperfine components $|FM_F\rangle$ of $|\psi_{X}^T \rangle$ in Eq.~\eqref{eq:trilo_nat1} are the average of the numerically obtained  natural orbitals for each $R$. However, the variation of these numerical coefficients around these average values is minimal. The values $\lambda_{X\pm}^T$ correspond to the eigenvalues of the reduced density matrix for the state $X$. Although a dependence on $R$ is written, these values also show negligible fluctuations relative to their average values [approximately (0.7,0.3) and (0.8,0.2)] and can be considered constant.

For other values of $n$ we find the same ground state atom spin components for the natural orbital. Therefore, as a first approximation,  trilobite states have hyperfine components with the same structure regardless of the value of $n$ and for all internuclear distances $R$. Additionally, the weight $\lambda^T_{X\pm}$ of each natural orbital also remains almost constant. In the perturbative approach presented in the following Section, the reason behind these particular values for the coefficients is identified. For the other $\Omega$ values we find a similar structure for the natural orbitals. The electronic contributions are given by the same functions $|u_{\pm} ^{(n)}(R) \rangle$ and the spin component has  approximately constant coefficients.

Next we study the natural orbitals for the butterfly states. In the spatial region where we can expect molecular binding the electronic component is formed by $n(\ell \geqslant \ell_{\mathrm{min}})$ states but now there is a small not always negligible contribution of $(n\pm 1)(\ell \geqslant \ell_{\mathrm{min}})$ states. For these PECs we find 2 different classes of states. First, those with pure $|m_j|=1/2$ and low$-\ell$ contribution, for which we can write
\begin{align}
|\psi_{X}^{B_1}\rangle \approx &\sqrt{\lambda^{B_1}_{X+}(R)}  \sum_{n}\gamma^{(n)}_{+}(R) |v_{+}^{(n)} (R)\rangle (\tilde{A}_X \, |1 \, 1 \rangle+\tilde{B}_X \, |2 \, 1 \rangle)+ \sqrt{\lambda^{B_1}_{X-}(R) } \sum_{n}\gamma_{-}^{(n)}(R) |v_{-}^{(n)} (R)\rangle \, |2 \, 2 \rangle  \nonumber \\
&+\sum_{j,F,M_F} b_X^{j,F,M_F}(R) |n^* p_{j} \, 3/2-M_F \rangle |F \, M_F \rangle,
\label{eq:but_nat1}
\end{align}
where the principal quantum number $n^*$ of the low-$\ell$ contribution is determined by the $p$ quantum defect. For $^{39}$K, $n^*=n+2$ as previously mentioned.

And second, those states with mixed projection $|m_j|=1/2,3/2$ that can be written as
\begin{align}
|\psi_{X}^{B_2}\rangle \approx& \sqrt{\lambda_{X3/2}^{B_2}(R)} \left( \sum_{n}\delta^{(n)}(R) |w_{3/2}^{(n)} (R)\rangle \right)(\bar{A}_X \, |1 \, 0 \rangle+\bar{B}_X \, |2 \, 0 \rangle) \nonumber \\
&+ \sqrt{\lambda_{X1/2}^{B_2}(R) } \sum_{n}\gamma_{n}(R) |w_{1/2}^{(n)} (R)\rangle \, (\bar{C}_X \, |1 \, 1 \rangle+\bar{D}_X \, |2 \, 1 \rangle),
\label{eq:but_nat2}
\end{align}

As for the trilobite case, here the eigenvalues $\lambda^{B_1}_{X\pm}$, $\lambda^{B_2}_{X1/2}$, $\lambda^{B_2}_{X3/2}$ and coefficients of the hyperfine component are practically constant over the spatial region of interest and independent of the principal quantum number $n$. We have used the notation $|v_{\pm}^{ (n)}\rangle$  and $|w_{1/2,3/2}^{ (n)}\rangle$ to highlight that these states have the same structure among themselves and the only parameter that differentiates them is $n$. The sum over $n$ includes terms with $n$ and $n\pm 1$. We make emphasis in the fact that the general structure is the same for all three classes of states: constant hyperfine components and reduced density matrix eigenvalues with an electronic state given by a linear combination of states with the same structure which only depend on $n$.

For butterfly LRRMs, the distinction between the two types of orbitals discussed in previous paragraphs is more evident for $^{39}$K than for $^{87}$Rb. Nevertheless the natural orbitals can always be written as a superposition of Eqs.~\eqref{eq:but_nat1}-\eqref{eq:but_nat2}.

\subsection{Perturbative model}
In the spin-independent description of LRRMs, perturbation theory provides a way of finding very good approximated analytic Rydberg electron wave functions and PECs \cite{Greene2000, Khuskivadze2002, Eiles2019}. Since states with low$-\ell$ have non-zero quantum defects with large non-integer parts they are energetically well differentiated.  As $\ell$ increases the quantum defects rapidly get smaller and are almost $j$-independent. For states with high$-\ell$, the splitting in the Rydberg energy is negligible compared to the strength of the Fermi pseudopotential and they can be treated as degenerate states for a given principal quantum number $n_0$. So first-order degenerate perturbation theory is used in this high$-\ell$ subspace. 

Here we extend the perturbative analysis presented in Refs. \cite{Greene2000, Eiles2019} to include spin effects. The first step is to define which part of the electronic Hamiltonian in Eq.~\eqref{eq:hspin} is identified as the unperturbed Hamiltonian. We take the Rydberg Hamiltonian along with the polarization potential as the unperturbed Hamiltonian, i.e., $\hat{H}_0=\hat{H}_{\mathrm{Ryd}}+\hat{H}_{\mathrm{pol}}$. Restricted to the set of nearly degenerate states with high angular momentum for a given principal quantum number $n_0$ we have approximately 
\begin{equation*}
\hat{H}_0 |n_0 (\ell \geqslant \ell_{\mathrm{min}}) j m_j \rangle |m_2 m_i \rangle=\epsilon_{n_0} |n_0 (\ell \geqslant \ell_{\mathrm{min}}) j m_j \rangle |m_2 m_i \rangle,
\end{equation*}
where
\begin{equation}
\epsilon_{n_0}=E^{\mathrm{Ryd}}_{n_0 \ell\geqslant \ell_{\mathrm{min}}}-\frac{\alpha_p}{2 R^4}.
\label{eq:edeg}
\end{equation}

To write Eq.~\eqref{eq:edeg} we have assumed that $E^{\mathrm{Ryd}}_{n_0 \ell\geqslant \ell_{\mathrm{min}} }$ does not depend on $\ell$ or $j$ for $\ell\geqslant \ell_{\mathrm{min}}$. This is a reasonable assumption since we are dealing with states of high angular momentum (quasi-degenerate hydrogen-like manifold) for which the energy splitting between states with different $\ell$ and $j$ is negligible compared to the Fermi pseudopotential. Under this assumption, the set of states $ \lbrace |n_0 \ell  j m_j \rangle |m_2 m_i \rangle \rbrace$, where $ \ell_{\mathrm{min}} \leqslant \ell \leqslant n_0-1$, $\ell-1/2 \leqslant j \leqslant \ell+1/2$, and the projections $(m_j; m_2,m_i)$ take all possible values, is a degenerate subspace of $\hat{H}_0$ for each internuclear distance $R$. We denote this subspace as $W_{n_0}$.

The system is approximately solved by first introducing the Fermi pseudopotential that fully includes spin-orbit coupling of the scattering process given by Eq.~\eqref{eq:Vfinal} as a perturbation to $\hat{H}_0$. Therefore  perturbation theory for degenerate states is applied. Additionally it is assumed that the coupling between states with different principal quantum numbers is negligible at this stage. To find the new eigenvectors and corresponding energies it is then necessary to diagonalize the matrix of $\hat{V}_{\mathrm{Fermi}}$ in the subspace $W_{n_0}$. Since the total Hamiltonian is block diagonal in $\Omega$, each value of $\Omega$ is considered separately.  Finally, it is assumed that the energy gap between $s$-wave and $p$-wave scattering is large enough to treat them separately. This assumption is supported by the results of the numerical calculation for values of the internuclear distance $R$ out of a small region around the avoided crossings. This means that away from these singular points a state can be identified either as trilobite or as butterfly.

The core idea of the perturbative method is to find functions $\widetilde{Q}_{LM_L}^{n \ell  j m_j}(R)$ from which the Fermi pseudopotential matrix $\mathbb{V}_{LS}$  for each $L$ and $S$ in a given $\Omega$ block, can be written in a simple way as a low-rank matrix with a structure similar to the spin independent case. In this situation, the eigenvectors are simply  a linear combination of $\widetilde{Q}_{LM_L}^{n \ell  j m_j}$ terms which are an extension of the $Q_{LM_L}^{n\ell j}(R)$ functions-- introduced in Ref.~\cite{Eiles2017} and exemplified in Section \ref{sec:theo} --that  take into account the angular momentum projection $m_j$. It results convenient to define the functions $\widetilde{Q}_{LM_L}^{n \ell  j m_j}(R)$  according to:
\begin{equation}
\widetilde{Q}_{LM_L}^{n \ell  j m_j}(R)=C_{\ell M_L, s_1 m_j-M_L}^{j m_j} Q_{LM_L}^{n \ell  j}(R).
\label{eq:Qbar}
\end{equation}
From these functions, we can construct a set of matrices $\mathbb{M}_k \in \mathcal{M}_{2(n_0-3) \times 2(n_0-3)}$ in the $W_{n_0}$ subspace that appear throughout the calculation for all values of $\Omega$. Generally, for a given $L$ each of these matrices can be written as 
\begin{eqnarray}
\left( \mathbb{M}_k \right)_{\ell j,\ell'j'}=\sum_{M_L,M_L'} \mathcal{B}_{M_L,M_L'}\widetilde{Q}_{LM_L}^{n_0 \ell j m_j} \widetilde{Q}_{L M_L'}^{n_0 \ell' j' m_j'}.
\label{eq:mkq}
\end{eqnarray} 

Up to this point the description is general. To proceed further we now make the assumption that there is no $^3P_J$ splitting. As we have pointed out, this is a valid enough approximation for LRRMs formed by $^{39}$K  and  lighter alkali atoms like Li and Na \cite{Eiles2018}. However, it breaks down for Rubidium and the consequences will be discussed below.  With the assumption of negligible $^3P_J$ splittings, the terms with mixed $M_L$ cancel each other and it is possible to separate $M_L=0$ and $M_L=\pm 1$ contributions. As a consequence, the $\mathbb{V}_{LS}$ matrices have a simple block structure in terms of $\mathbb{M}_k$ with well-defined $M_L$. It is straightforward to find their eigenvalues and eigenvectors. The eigenvalues  are then $J$-independent and and have generally  the form
\begin{equation}
\lambda_{LS}(R)= 2 \pi(2L+1) a(L,S,J,k(R)) \sigma_{LM_L}^{n_0 m_j}(R),
\label{eq:llsj}
\end{equation}
where the normalization constants are defined as
\begin{equation}
\label{eq:sigmanorm}
\sigma^{n m_j}_{L M_L}(R)=\sum_{\ell j} \left| \widetilde{Q}_{L M_L}^{n\ell jm_j}(R)  \right|^2.
\end{equation}
 
The corresponding eigenvectors can be written as 
\begin{equation}
|v_{LM_L}^{(n_0)}\rangle=\sum_{m_2,m_i} D_{m_2,m_i} |\alpha_{\Omega-m_2-m_i,L M_L}^{(n_0)} \rangle |m_2 \, m_i \rangle,
\label{eq:vpert}
\end{equation}
where the fundamental electronic states are
\begin{equation}
| \alpha_{m_j,L M_L}^{(n)} \rangle =\frac{1}{\sqrt{\sigma^{n m_j}_{L M_L}(R)}} \sum_{\ell,j} \widetilde{Q}_{L M_L}^{n\ell jm_j}(R) |n \ell  j m_j \rangle.
\label{eq:edo_alfa}
\end{equation}
Before continuing, it is important to make two observations. The first is that the coefficients  $D_{m_2,m_i}$ are numbers that do not depend on $n$ or $R$. And the second is that although the Rydberg electron component of the eigenstates (given by $| \alpha_{m_j,L M_L} \rangle$) could be thought of as similar to those obtained in the spin-independent description, our perturbative method provides the correct multiplicity of states the non-trivial values of the coefficients $D_{m_2,m_i}$ that make the full spatial-spin state $|v_{LM_L}^{(n_0)}\rangle$ an  approximate  eigenvector of the spin-orbit dependent Fermi pseudopotential.
More details of the method are presented in Appendix~\ref{section:Fermi}  where, as an illustrative example, explicit values of these coefficients  for $\Omega=3/2$ are presented in Eqs.~\eqref{eq:trilo_psing} and \eqref{eq:vecfermi}  for trilobite states and Eqs. \eqref{eq:vecfermiP}-\eqref{eq:vecfermiP2} for buttterfly states.

So far we have not considered the hyperfine interaction of the ground state atom $\hat{H}_{\mathrm{HF}}$. We include this term as an additive perturbation to the Hamiltonian $\hat{H}_0+\hat{V}_{\mathrm{Fermi}}$ for which we have approximate solutions $|v_{LM_L}^{(n_0)}\rangle$ given by Eq.~\eqref{eq:vpert}.
Depending on $\Omega$, the eigenvectors for Fermi singlet or triplet pseudopotential have degeneracy $g_{LS}$, so that perturbation theory of degenerate states within the corresponding $g_{LS}$ dimensional subspace must be used. By diagonalizing the hyperfine interaction in this subspace, the degeneracy is completely broken (in almost every case) and different trilobite (butterfly) states are found for $L=0 \, (1)$. This procedure is described in general and exemplified in detail for $\Omega=3/2$ in the Appendix \ref{section:hyperfine}. For the $\Omega=3/2$ example, the explicit expressions for the approximate eigenvectors of the full Hamiltonian including Fermi and hyperfine terms are given by Eqs. \eqref{eq:T1}-\eqref{eq:T2} and \eqref{eq:B1}-\eqref{eq:B6} in that Appendix.

For each value of $n_0$, perturbation theory predicts $g_{LS}$ different states for $L$, $S$ Fermi pseudopotential interaction. For example, for $\Omega=3/2$ we have $g_{00}=1$ and $g_{01}=2$ for trilobite states. This prediction is consistent with the result of numerical diagonalization shown in Fig.~\ref{fig:Ktrilospin}. 

\begin{figure}[ht!]
	\centering
	\subfloat{%
		\centering
		\includegraphics[width=0.325\textwidth]{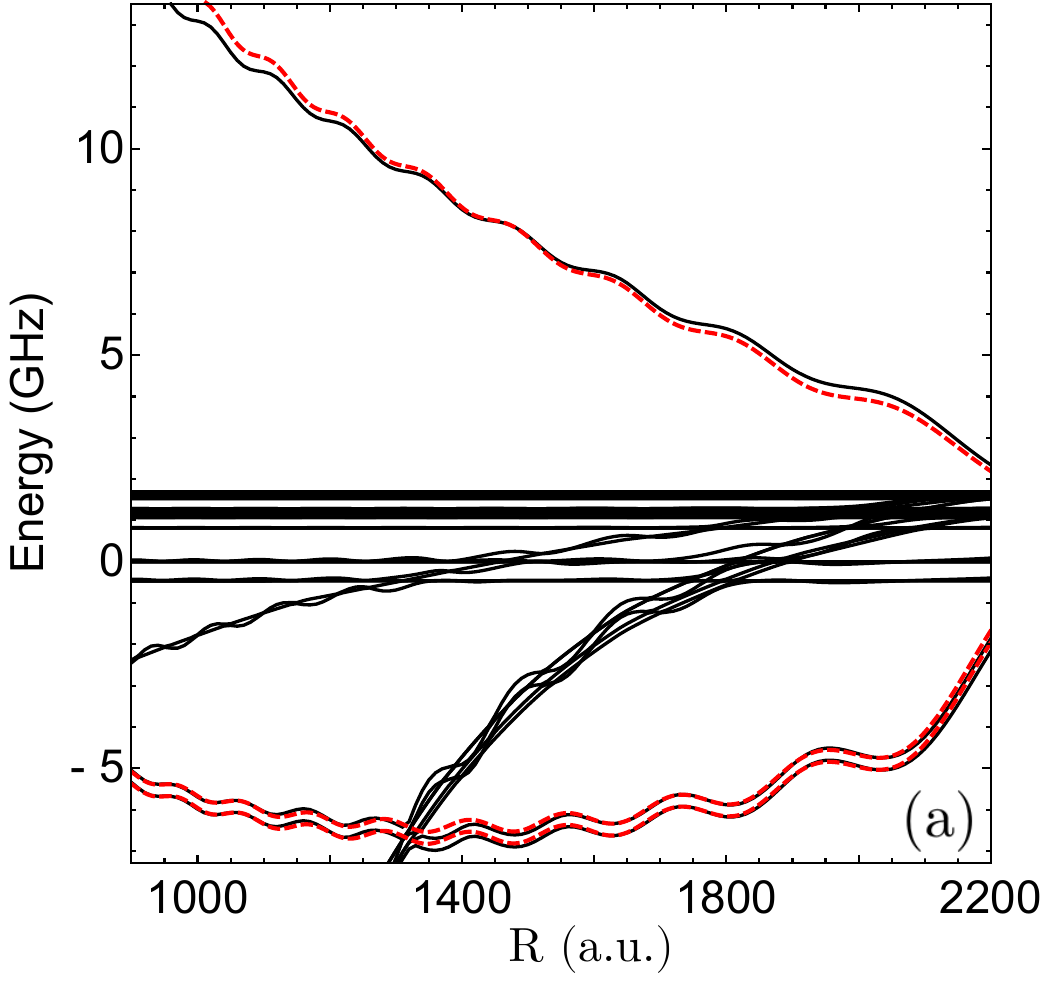}
		\label{fig:compara}
	}\hfil
	\subfloat{%
		\centering
		\includegraphics[width=0.315\textwidth]{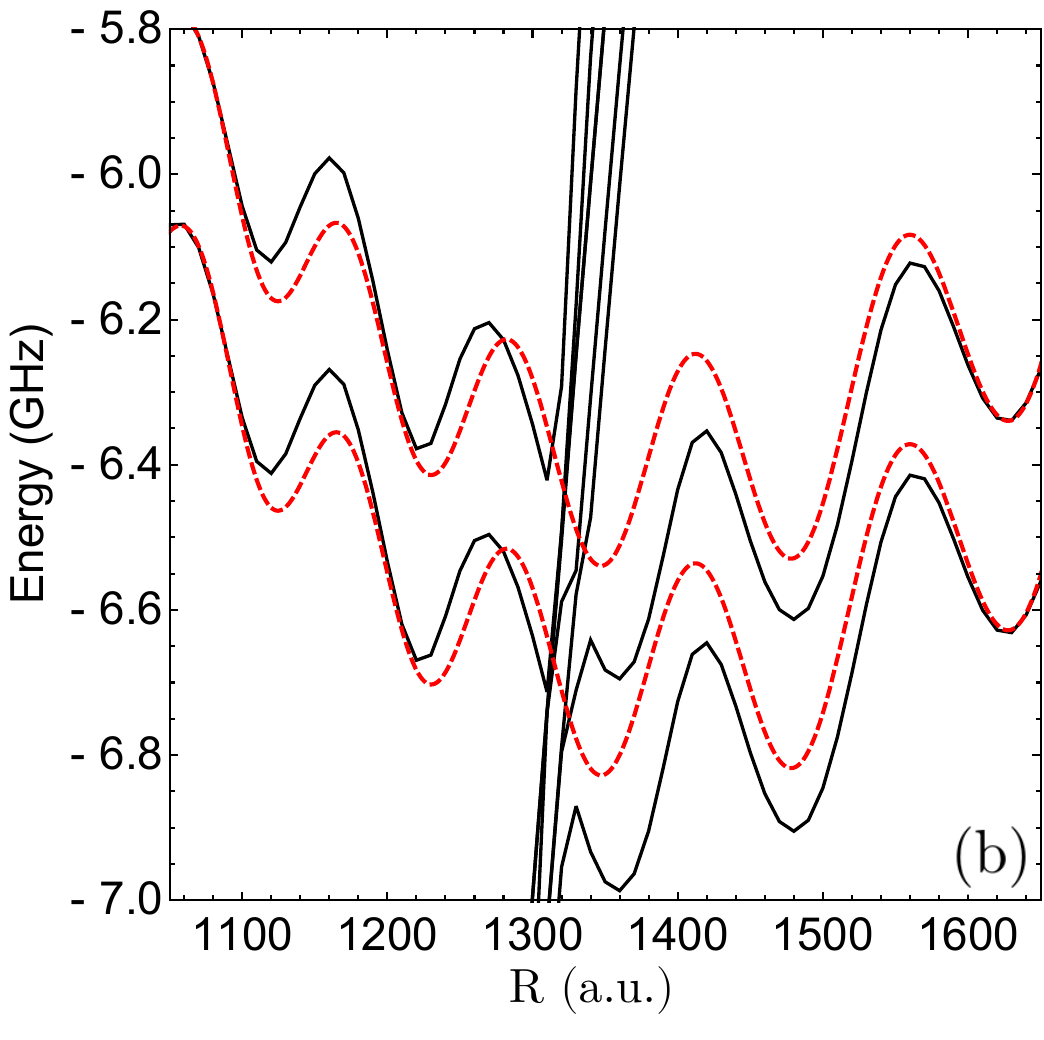}
		\label{fig:compara2}
	}\hfil
	\subfloat{%
		\centering
		\includegraphics[width=0.325\columnwidth]{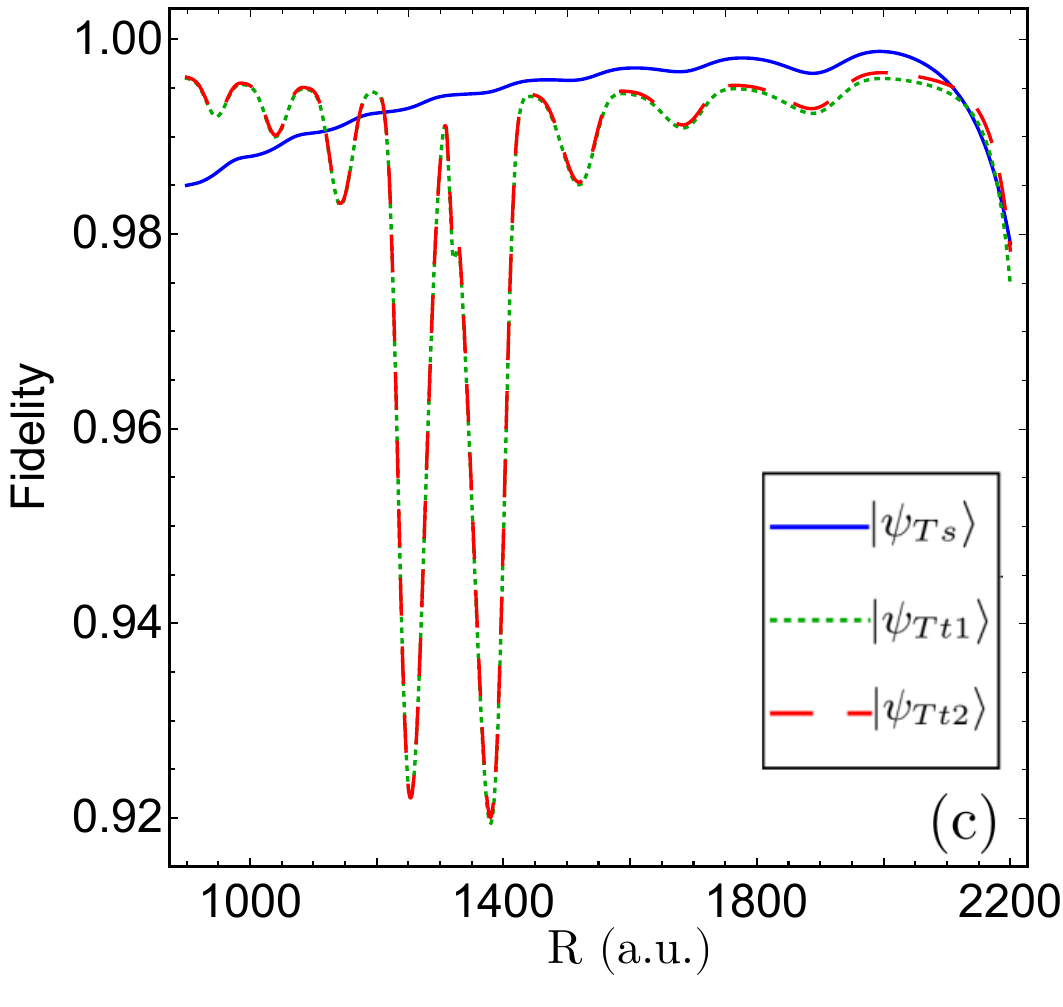}
		\label{fig:fidelidad}
	}	
	\caption{ Comparison between numerical trilobite PECs as a function of the nuclear distance $R$ from complete diagonalization (solid line) and those predicted by the perturbative analysis (red dotted line) for $\Omega=3/2$ and $n_0=34$ in $^{39}$K molecules.(a) depicts  $900 a_0 \le R\le 2200 a_0$, while the region around avoided crossings is highlighted in (b). (c) Fidelity between numerical and perturbative singlet $|\psi_{Ts}\rangle$ and triplet $|\psi_{Tt}\rangle$ states. }
\end{figure}

Figure~\ref{fig:compara} shows a comparison between the energy of the numerical diagonalization described in Subsection~III.A and the prediction of the perturbative model for singlet and triplet trilobite states for  $^{39}$K and $\Omega=3/2$ . It can be seen that in general, perturbative analysis produces excellent results for both singlet and triplet terms. For the triplet PECs, the differences between the energies are mainly due to the contribution of $p-$wave scattering. The shape resonance causes a slight shift of the energy levels in the neighborhood of the avoided crossings and such effect is not included in the perturbative model.

For trilobite molecules, the mixture of different hyperfine states is mostly caused by the hyperfine interaction. The perturbative trilobite eigenstates represent an excellent approximation to those obtained numerically. The quantum fidelity between the numerical and perturbative states is a simple way to quantify how good the approximation is. Fig.~\ref{fig:fidelidad} shows the fidelity as a function of the internuclear distance. A value above 90 \% for all values of $R$ in the region of interest is achieved. This value is even greater outside the avoided crossing regions. 

Similar results are found for the other possible values of $\Omega$ and $L=1$ scattering. The Fermi pseudopotential matrix can always be written as a block matrix formed by matrices $\mathbb{M}_k$ with an $n,\ell,j$ independent prefactor that depends only on the projections for the corresponding block and the scattering channel. It is important to note that the only dependence on $n$ in trilobite states is through the electronic state $|\alpha_{\pm {\scriptstyle \frac{1}{2}}, 0 \, 0}^{(n)} \rangle$. The ground atom component of the state is always the same for all $n$.

In a consistent way with numerical natural orbitals for $^{39}$K, we found two classes of perturbative butterfly states. The $M_L=0$ states that correspond to $\Sigma$ molecular symmetry are denoted as radial butterflies and written explicitly as Eqs. \eqref{eq:B1}-\eqref{eq:B2} in the Appendix below. And the $|M_L|=1$ angular butterfly states of Eqs.~\eqref{eq:B3}-\eqref{eq:B6} with $\Pi$ symmetry. The numerical natural orbitals are a linear combination of these perturbative butterfly states for $n-1,n,n+1$ manifolds with a low$-\ell$ contribution. The explicit expressions are Eqs.~\eqref{eq:mixn}-\eqref{eq:mixn2} of the Appendix~\ref{section:Fermi}. 

As mentioned before, in $^{87}$Rb there is a considerable $^3P_J$ splitting of the phase shifts. This prevents the $\mathbb{V}_{LS}$ matrices from having a simple separable block structure. This can be considered the source of the difficulty in describing the butterfly dominated LRRMs for $^{87}$Rb because it is not possible to separate the contributions of different $M_L$. 

\section{Total spin quasi-symmetry for high$-\ell$ LRRMs}
Now, with compact expressions for the high$-\ell$ states for $s-$wave and $p-$wave scattering at hand, we are interested in studying their possible symmetries.  In this Section we present an analysis in order to show that the molecule total spin $\vec{N}=\vec{s}_1+\vec{s}_2+\vec{i}$ results in good quantum numbers that identify each of the spin-orbitals used to describe high$-\ell$ molecular states and PECs in determined spatial regions. Although the quantum number $N$ has been used to some degree in the description of LRRMs \cite{Deiss2020}, its quality as an operator associated with a symmetry has not been explored.

We can write the total spin $\vec{N}$ by prioritizing the hyperfine coupling $s_2-i$, as this way we have obtained the perturbative states. The operator is written as $\vec{N}=\vec{s}_1+\vec{F}$ for this coupling scheme, and the corresponding vectors are given by
\begin{equation}
|(s_1F) \, N, \, M_N  \rangle =\sum_{m_1,M_F} C_{s_1 \, m_1, \, F \, M_F}^{N \, M_N} |s_1 m_1 \rangle_{s_1} |F M_F\rangle_F
\label{eq:N1}
\end{equation}
where $M_N$ is the projection of $\vec{N}$ onto the internuclear axis. 

However, taking into account the Fermi pseudopotential, if the $s_1-s_2$ coupling is first performed $\vec{N}=\vec{S}+\vec{i}$ the states are determined by 
\begin{equation}
|(S i) \, N, \, M_N  \rangle =\sum_{M_S, m_i} C_{SM_S,im_i}^{N M_N} |S M_S\rangle_S |i m_i \rangle_i
\label{eq:N2}
\end{equation}

To express these well-defined $(N;S,i)$ states using hyperfine coupling, it is necessary to carry out the base transformation associated with the change in coupling type. This is commonly done in the study of atomic spectra that is outside the LS or $j-j$ coupling regime, so that one type of coupling may be more suitable than another for different configurations \cite{Racah1943, Cowan1965}. Explicitly, for a given $N$ and starting from a $s_1-s_2$ coupling, the transformation between representations is given in terms of 6-$j$ symbols, 
\begin{align}
|(S i) \, N, \, M_N \rangle = \sum_{F=|i-s_2|}^{i+s_2} (-1)^{i+s_1+s_2+N} \sqrt{(2F+1)(2S+1)} 
\begin{Bmatrix}
F & s_1 & N\\
S & i & s_2
\end{Bmatrix}
|(s_1F) \, N, \, M_N  \rangle 
\label{eq:racah}
\end{align}

Now, we shall show that the perturbative states for all different values of $\Omega$ found in previous Sections result to have $N$ as a good quantum number within the $s_1-s_2$ coupling scheme. First, we express the electronic state with the spin of the Rydberg electron written explicitly. Starting from Eq.~\eqref{eq:edo_alfa}, whenever the $j$ dependence of the high$-\ell$ quantum defects is neglected, the electronic orbitals can be approximately written as 
\begin{equation}
| \alpha_{m_j,L M_L}^{(n)} \rangle \approx \frac{1}{\sqrt{\sigma_{LM_L}^n(R)}}\left[ \sum_{\ell \geqslant \ell_{\mathrm{min}}} Q_{LM_L}^{n \ell}(R) |n \ell M_L \rangle \right] |s_1 \, m_j-M_L  \rangle_{s_1}:= | \Theta_{LM_L}^{(n)} \rangle |s_1 \, m_j-M_L  \rangle_{s_1}, 
\label{eq:alfa2}
\end{equation} 
where  $| \Theta_{LM_L}^{(n)} \rangle$ are the spin-independent trilobite and butterfly orbitals \cite{Eiles2019}. Substituting Eq.~\eqref{eq:alfa2} in the expressions for the perturbative states, we find molecular states in which the spatial and spin  degrees of freedom are completely separated. When writing the resulting states according to Eq.~\eqref{eq:N1} we find that the weight of each $F$ contribution is precisely the coefficient that  appears in Eq.~\eqref{eq:racah} and the dependence on $N$ is made clear. It results that  all perturbative states are of the form  
\begin{equation}
|\Psi \rangle:=|\Theta_{LM_L}^{(n)}(R) \rangle |(Si) \, N, \, \Omega-M_L \rangle.
\label{eq:psiNcompleto}
\end{equation}
As a consequence, each high-$\ell$ perturbative state can be labeled by the total angular momentum projection $\Omega$, the principal quantum number $n$, the total molecular spin $N$ and the scattering channel  they come from $L, M_L$ and $S$. We denote this set of quantum number as $\xi:=(n,L,M_L,S,N;\Omega)$. The projection of $\vec{N}$ is not an independent observable since $M_N=\Omega-M_L$. We note that all spatial dependence is contained in the $|\Theta \rangle$ orbitals and that the spin component is independent of $L$. 

Through this general expression for $|\Psi_{\xi} \rangle$, it is possible to determine the number of perturbative states that exist for each scattering channel for a given value of $\Omega$. Remembering that for $^{39}$K we have a nuclear spin $i=3/2$, the possible values of $N$ are $N=1/2,\ 3/2,\ 5/2$. For our working example $\Omega=3/2$, and given that only the scattering by the two first partial waves ($|M_L|\leqslant1$) is considered, there are two well-defined $N$ states for $S=0$, 
\begin{equation}
|(0, \,3/2) \, 3/2, \, M_N  \rangle, \hspace{0.8cm} 1/2 \leqslant M_N \leqslant 3/2 .
\end{equation}
And six states for $S=1$
\begin{align}
|(1, \,3/2) \, 1/2, \, 1/2  \rangle,& \\
|(1, \,3/2) \, 3/2, \, M_N \rangle,& \hspace{0.8cm} 1/2 \leqslant M_N \leqslant 3/2 , \\
|(1, \,3/2) \,5/2, \, M_N \rangle, &\hspace{0.8cm} 1/2 \leqslant M_N \leqslant 5/2 ,
\end{align}

Taking into account the values of $L$ compatible with each of these 8 spin states, there are a total of 11 states. This is exactly the number of states obtained with the perturbative analysis.

\begin{figure}[ht!]
	\centering	
	\includegraphics[width=0.93\columnwidth]{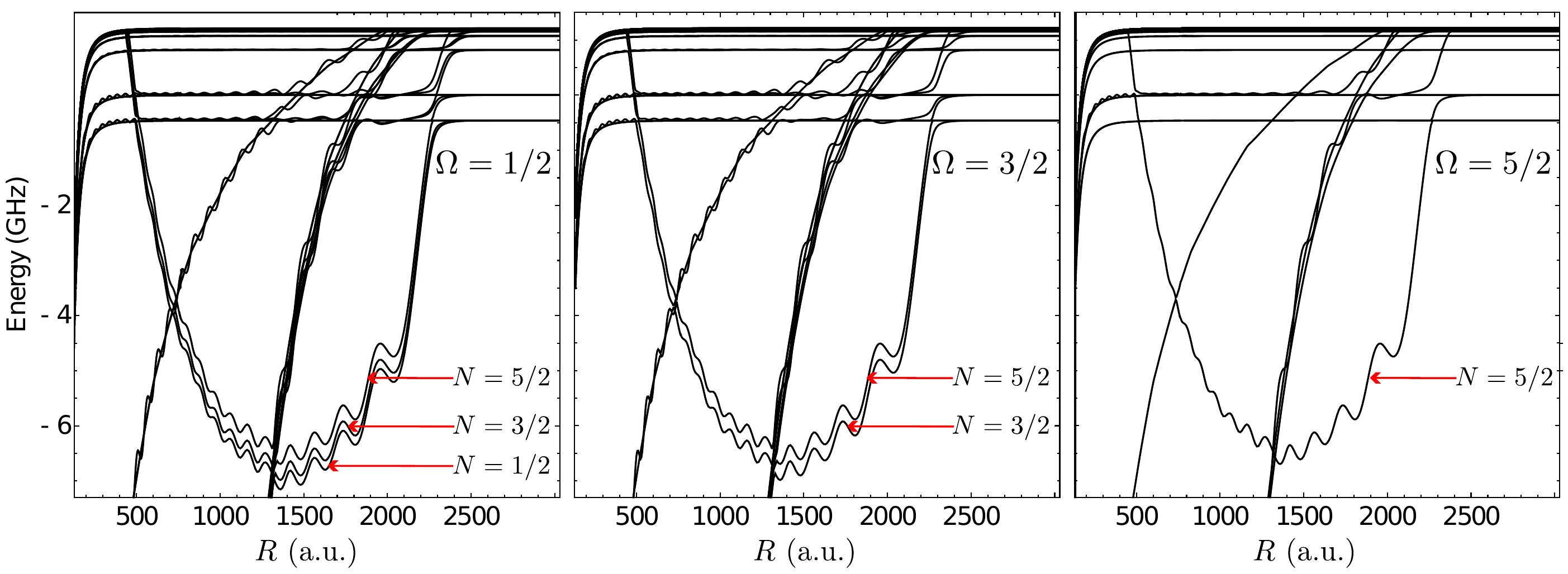}
	\caption{Values of the good quantum number $N$ for the $^{39}$K LRRM triplet trilobite PECs already illustrated in Fig.~\ref{fig:Ktrilospin}.}	
	\label{fig:trilosN}
\end{figure}

As it has been shown in the previous Section, the perturbative trilobite states are practically equal to the numerical eigenvectors. As a consequence, the symmetry generated by $\vec{N}$ is present in the states resulting from the complete diagonalization of the Hamiltonian. Figure~\ref{fig:trilosN} illustrates the triplet trilobite PECs and the corresponding $N$ values. 

For $^{39}$K, since the spin component is independent of the principal quantum number $n$, a linear combination of butterflies of the same type has the same value of $N$ as the elements of the linear combination. For different values of $\Omega$ and within a range of principal quantum numbers around $n=34$, we performed a careful analysis to verify that the coefficients $G_{j,M_F}^{(F)}(R)$ of  Eq.~\eqref{eq:mixn} are such that the low angular momentum contribution of the state also has a well-defined value of $N$ on the region of interest, which coincides with the value of $N$ for the butterfly character of the state. Therefore, we verify that under the used approximation of taking all $^3P_J$ phase shifts to be equal, the symmetry given by $N$ is also present for the numerical butterfly states for this atomic species . Accordingly, for $^{39}$K the triplet dominated butterflies PECs (and eigenstates) asymptotically correlated to one $n$ in an $\Omega$ block can be labeled by $(N,M_L)$ similar to Fig. \ref{fig:trilosN}. However, we have to keep in mind that there are two radial ($M_L=0$) butterfly PECs for each value of $N$ as a consequence of the splitting when including low-$\ell$ states. This produces two sets of radial butterfly PECs, and in each of said sets all the $N$ values compatibles with $\Omega$ are present. Consider for example $\Omega=3/2$, the four angular butterflies correspond to the $(N,M_L)$ values $(1/2,1)$, $(3/2,1)$,$(5/2,1)$,$(5/2,-1)$ in ascending order of energy. We note that the two $N=5/2$ PECs are degenerate. In each doublet of radial butterfly PECs, the lower curve corresponds to $N=3/2$ and the higher to $N=5/2$.

Since the symmetry given by $\vec{N}$ is an approximate one, it results necessary to quantify how approximate this symmetry is. To this end, we have calculated the expectation value of $\hat{N}^2$ using its spectral decomposition given by
\begin{equation}
\hat{N}^2=\vec {N} \cdot \vec {N}=\sum_{N,M_N,S} N(N+1) \, \hat{P}_{N,M_N;S},
\end{equation}
where $\hat{P}_{N,M_N;S}$ is a projector onto the state with well-defined $N$ and $M_N$ restricted to electronic spin $S$. Note that we are using the $S-i$ coupling for $N$ and we do not write explicitly $i$ neither for $^{39}$K nor $^{87}$Rb as it has a fixed value $i=3/2$. We sum over spin singlet and triplet contributions as we are interested in the symmetry given only by $\vec{N}$.

\begin{figure}[ht!]
	\centering	
	\subfloat{%
		\includegraphics[width=0.4\columnwidth]{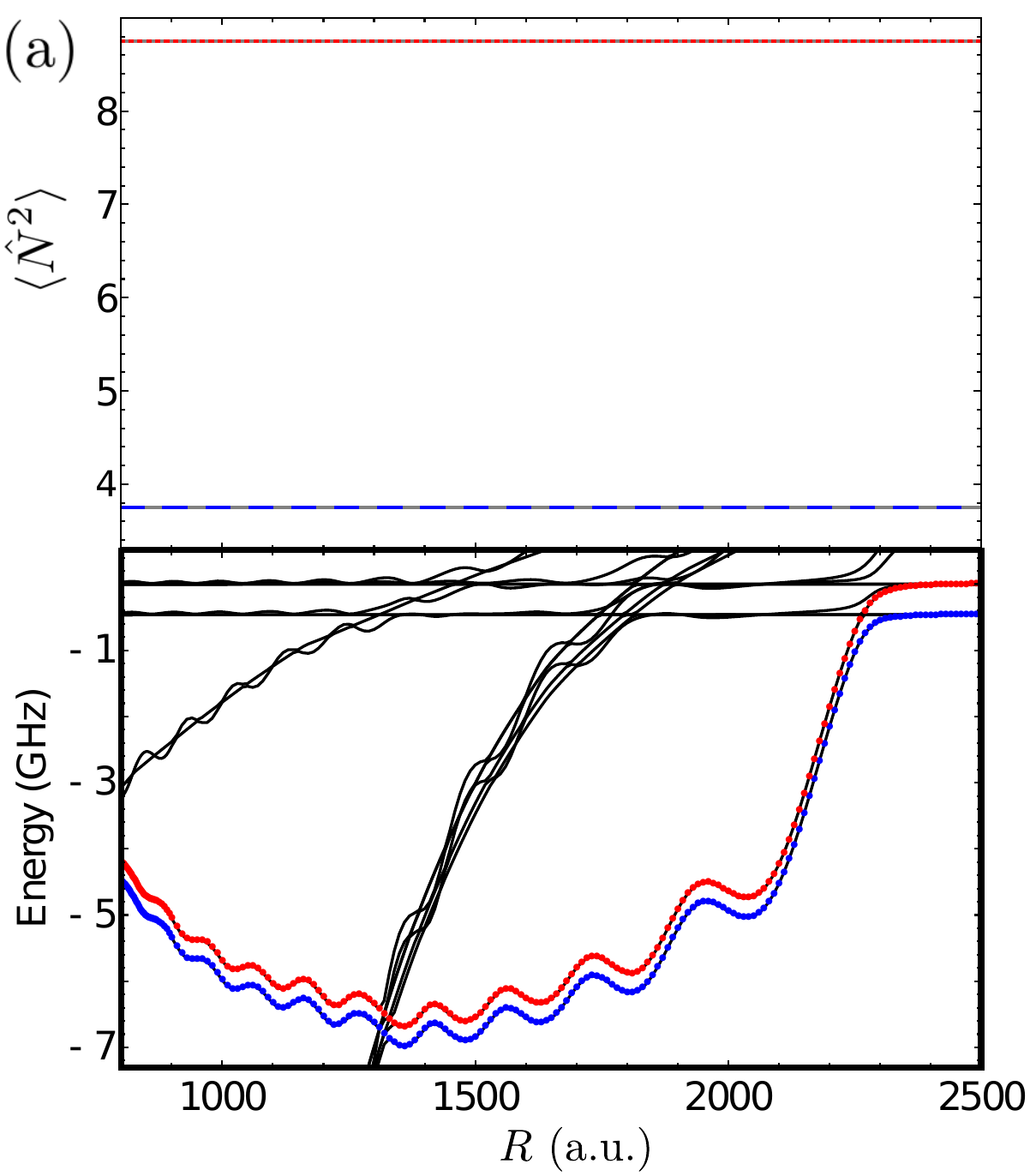}
	} \hspace{0.1cm}
	\subfloat{%
		\includegraphics[width=0.4\columnwidth]{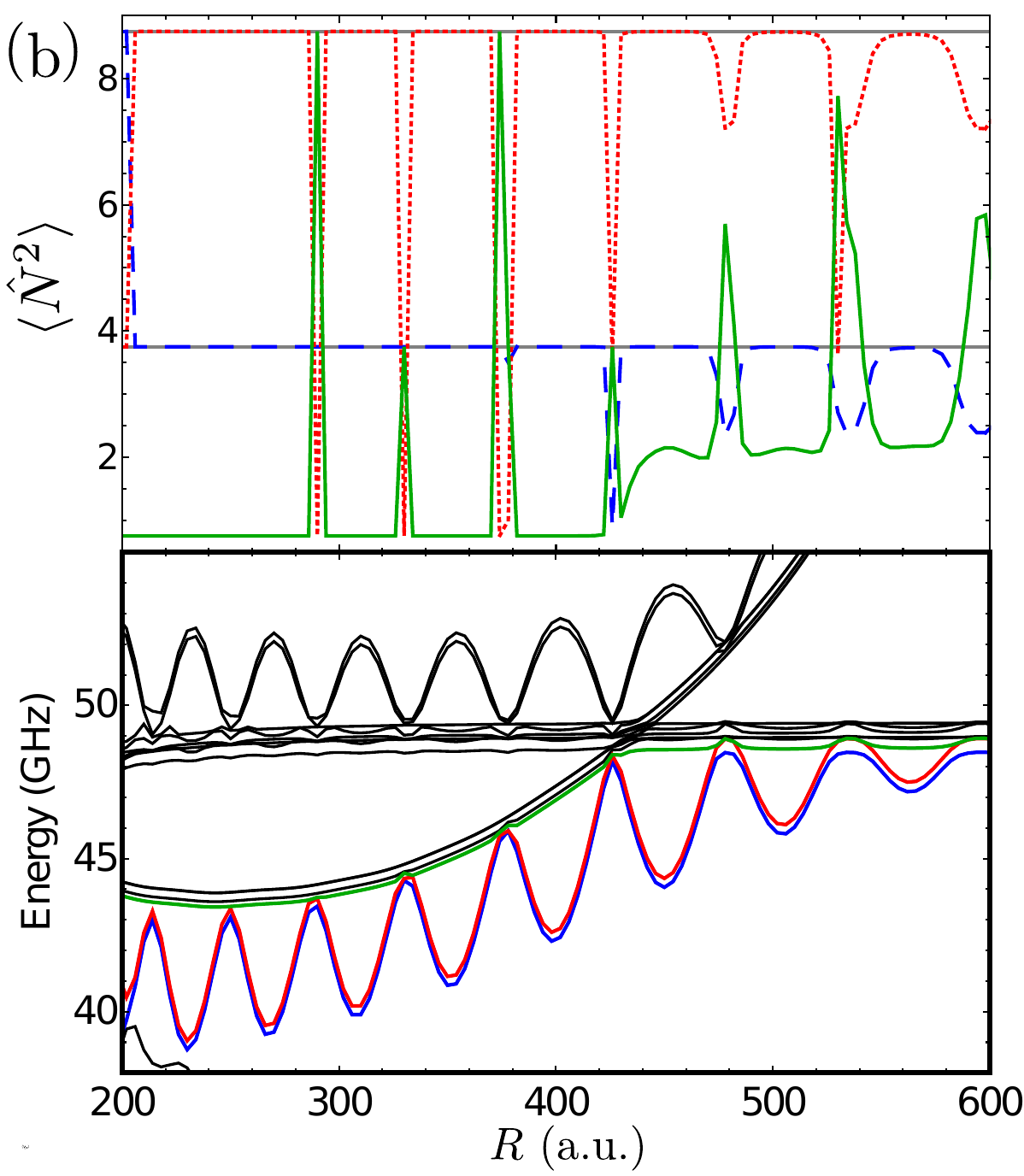} 
	}	
	\caption{Expectation value $\langle \psi|\hat{N}^2|\psi \rangle$ and corresponding PECs as a function of the internuclear distance $R$ for the numerical (a) trilobite and (b) butterfly states with $\Omega=3/2$ in $^{39}$K. For the trilobite curves, blue dashed corresponds to $N=3/2$ and red dotted to $N=5/2$ states.
	For butterfly curves blue dashed, red dotted and green continuous correspond to the lower $N=3/2$ and $N=5/2$ radial, and $N=1/2$ angular states respectively. The values corresponding to the exact $N(N+1)$ are shown in faded gray lines. The energy is relative to the $|34 \, f_{5/2} \rangle |2 \rangle$ state.}	
	\label{fig:NK}
\end{figure}

We first consider the case of $^{39}$K LRRMs. Figure~\ref{fig:NK} illustrates the results for the two trilobite states and the lower $N=3/2,5/2$ radial and $N=1/2$ angular butterfly states with $\Omega=3/2$ for different values of the internuclear distance $R$. For trilobite molecules we find a practically constant value corresponding to the well-defined values of $N$ shown in Fig. \ref{fig:trilosN}. For the case of butterfly molecules and in the region where bonded states exist ($R \lessapprox 420 \, a_0$) we find that $\langle \hat{N^2} \rangle$ is almost constant and deviates of the associated $N(N+1)$ value only on the avoided crossings between the PECs. For $R \gtrapprox 420 \, a_0$ the expectation value starts to show a real variation as a function of the internuclear distance and as a consequence $N$ is no longer a good quantum number in this region. 

\begin{figure}[ht!]
	\centering	
	\subfloat{%
		\includegraphics[width=0.4\columnwidth]{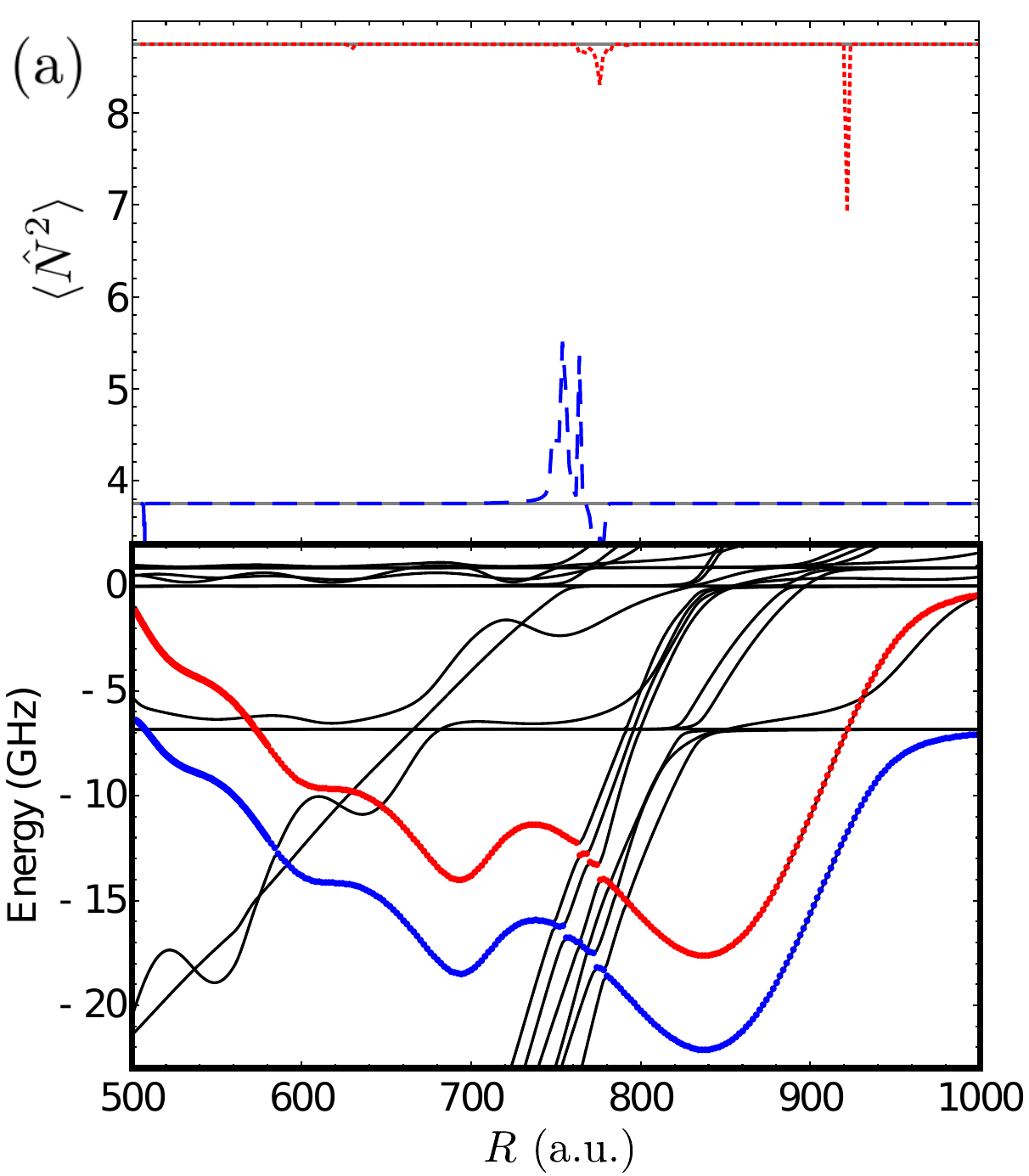}
	} \hspace{0.1cm}
	\subfloat{%
		\includegraphics[width=0.4\columnwidth]{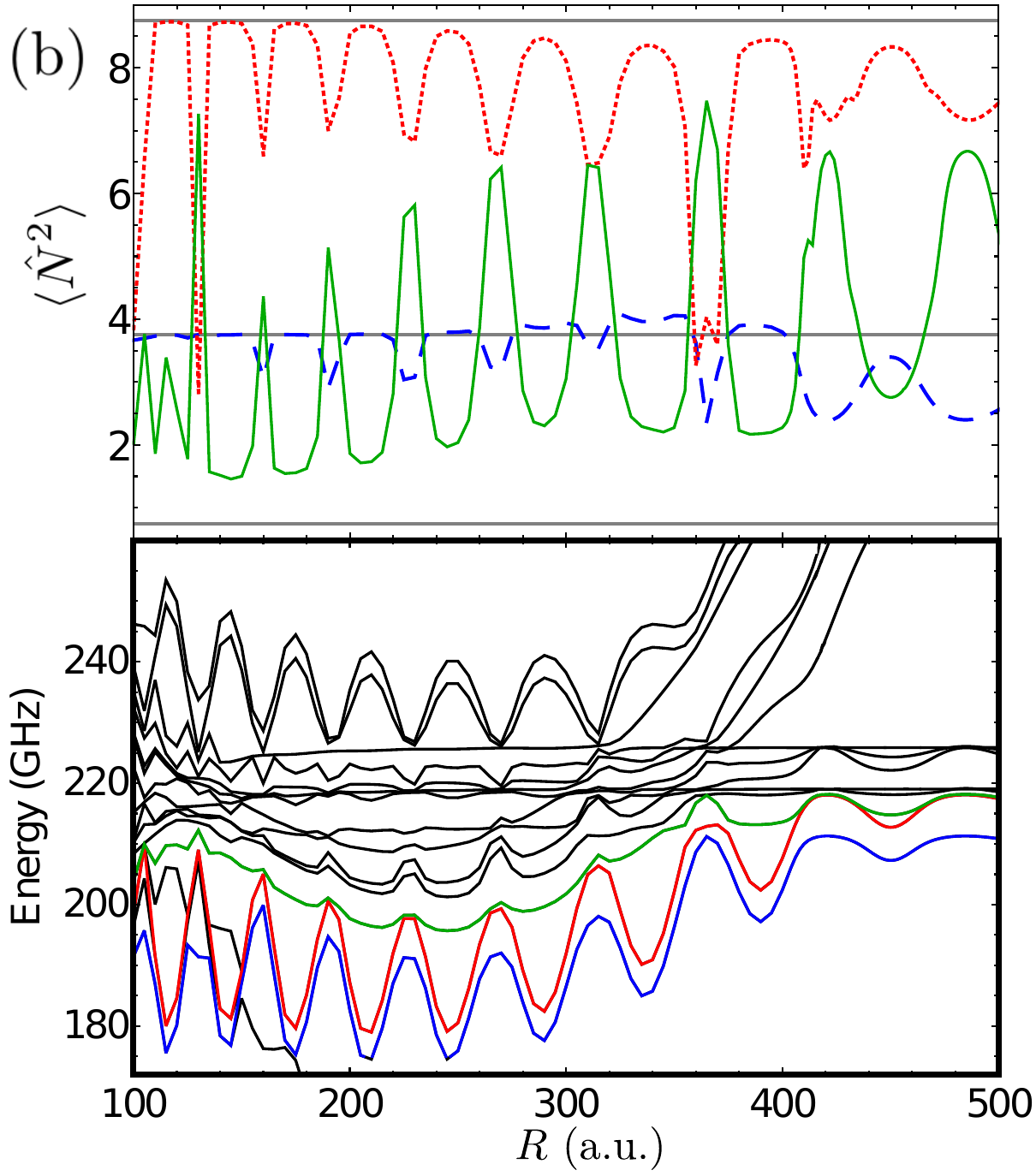} 
	}	
	\caption{Expectation value $\langle \psi|\hat{N}^2|\psi \rangle$ and corresponding PECs as a function of the internuclear distance $R$ for the numerical (a) trilobite and (b) butterfly states with $\Omega=3/2$ in $^{87}$Rb. For the trilobite curves, blue dashed corresponds to $N=3/2$ and red dotted to $N=5/2$ states. For the butterfly curves blue dashed, red dotted and green continuous correspond to the lower $N=3/2$ and $N=5/2$ radial, and $N=1/2$ angular dominated states respectively. The values corresponding to the exact $N(N+1)$ are shown in faded gray lines. The energy is relative to the $|22\, f_{5/2} \rangle |2 \rangle$ state.}	
	\label{fig:NRb}
\end{figure}
A similar analysis has been performed for $^{87}$Rb LRRMs. For this system, there is experimental evidence for spin-orbit interaction~\cite{Deiss2020} and  the role of $\vec N$ must be carefully studied. In this case,  the $^3P_J$ splittings are not negligible which enhances the role of scattering coupling scheme $L$-$S$ in the molecule formation. This splitting affects the performance of $\vec{N}$ as an approximate symmetry. In Figure~\ref{fig:NRb} we present the expectation value of $\hat{ N}^2$ for Rb$_2$ molecules for some representative PECs. As the effect of $p-$wave Fermi interaction on trilobite states is minimal we find that away of the avoided crossings the $N$-symmetry is also present in the general trilobite case. On the other hand, for butterfly states the symmetry is broken. The label $(N,M_L)$ can no longer be used to identify each PEC. As it is illustrated in Figs.~\ref{section:Nproj}-\ref{fig:projRb} in the Appendix, each PEC have contributions of all three $N$ values. Nevertheless, the term $(N,M_L)$ with dominant contribution in any PEC correspond to the expected state in absence of $^3P_J$ splitting. To perform a reasonable description  of the states associated with butterfly molecules when there is $^{3}P_J$ splitting, it is necessary to include states with different values of $N$ and $M_L$.

From the two examples presented here it can be observed that the greatest deficiencies in $N$ as a good quantum number occur when two (or more) PECs are close to each other. Even in the case of $^{39}$K, in Fig. \ref{fig:NK} the points where the value of $N$ changes suddenly are precisely in the avoided crossings between two PECs, that is when they are almost degenerate. A similar behavior is observed in $^{87}$Rb, for in Fig. \ref{fig:NRb} the peaks that are furthest from the value corresponding to a well-defined $N$ are located  in the internuclear separations in which the PECs become almost degenerate. This suggests that degeneration between PECs is one of the factors contributing to the $N$-symmetry breaking.

Despite the $N-$symmetry breaking, the characterization of the states in terms of superposition of spin-orbitals that include quantum number provides a compact expression of the electronic wavefunction: for this case where $i=3/2$ there are only three possible $N$ values. This is studied in detail in the following Section.

\section{Spin-Orbital compact basis}
According to the analysis presented in the previous Section, to describe the numerical eigenstates of the complete Hamiltonian, it is sufficient to use at most a set of perturbative states with 2 or 3 values of principal quantum number $n$ and some states with low angular momentum. Returning to our example of $\Omega=3/2$, this translates to approximately 60 states needed to describe the set of resulting PECs and eigenvectors in the region of interest. This number is significantly smaller than the 2300 elements of the Rydberg state basis. 

The set  $\lbrace |\Psi_{\xi} \rangle \rbrace$ of states given by Eq.~\eqref{eq:psiNcompleto} constitutes a compact set, which is convenient to use for the description of high$-\ell$ LRRMs. On one hand, the numerical advantage is observed with the significantly smaller size of the basis. On the other hand, analytically it allows for a better understanding of each contribution of the Hamiltonian and the interactions that are relevant between different classes of states.  Since the spin components of $|\Psi_{\xi} \rangle$ are orthogonal the matrix elements of the Rydberg and polarization interactions are diagonal in spin and proportional to the spatial matrix element. As the $|\Theta_{LM}^{(n)}\rangle$ are in general not orthogonal, this spatial matrix element is given simply by the trilobite overlap matrix \cite{Eiles2019} 
\begin{equation}
\boldsymbol{\Upsilon}^{LM_L}_{L'M_L'}(n,R)= \sum_{\ell \geqslant \ell_{\mathrm{min}}}^{n-1} Q^{n \ell}_{LM_L}(R) Q^{n  \ell}_{L'M_L'}(R). 
\end{equation}

Using the fact that the normalization constant $\sigma_{LM_L}^n$ can be recast in terms of the overlap matrix, for our high$-\ell$ states we have the orthogonality condition 
\begin{equation}
\langle \xi | \xi' \rangle= \delta_{nn'} \delta_{M_L M_L' } \delta_{S S'} \delta_{NN'} \frac{\boldsymbol{\Upsilon}^{LM_L}_{L'M_L'}(n,R)}{\sqrt{\boldsymbol{\Upsilon}^{LM_L}_{LM_L}(n,R)\boldsymbol{\Upsilon}^{L'M_L}_{L'M_L}(n,R)}}.
\end{equation}

For a given $\Omega$ block the matrix elements of $\hat{H}_0$ are
\begin{equation}
\langle \Psi_{\xi}| \hat{H}_{\mathrm{Ryd}}+\hat{H}_{\mathrm{pol}} | \Psi_{\xi'} \rangle=\delta_{S S'} \, \delta_{N N'}  \,  \delta_{M_L M_L' } \,  \delta_{nn'} \, \frac{\boldsymbol{\Upsilon}^{LM_L}_{L'M_L'}(n,R)}{\sqrt{\boldsymbol{\Upsilon}^{LM_L}_{LM_L}(n,R) \, \boldsymbol{\Upsilon}^{L'M_L'}_{L'M_L'}(n,R)}} \left( E^{\mathrm{Ryd}}_{n \ell\geqslant \ell_{\mathrm{min}}}-\frac{\alpha_p}{2 R^4} \right).
\end{equation}

For the hyperfine term  we use Eq.~\eqref{eq:racah} to write the states in the $|FM_F\rangle_F$ basis. We obtain
\begin{align}
\langle \Psi_{\xi}| \hat{H}_{\mathrm{HF}} | \Psi_{\xi'} \rangle=&\frac{A_{\mathrm{HF}}}{2} \delta_{nn'} \, \delta_{M_LM_L'} \, \frac{\boldsymbol{\Upsilon}^{LM_L}_{L'M_L'}(n,R)}{\sqrt{\boldsymbol{\Upsilon}^{LM_L}_{LM_L}(n,R) \, \boldsymbol{\Upsilon}^{L'M_L'}_{L'M_L'}(n,R)}} \delta_{N N'} (-1)^{2i+2N} \sqrt{(2S+1)(2S'+1)}  \nonumber \\
&\sum_{F=|i-s_2|}^{i+s_2} (2F+1)[F(F+1)-i(i+1)-s_2(s_2+1)] 
\begin{Bmatrix}
F & s_1 & N\\
S & i & s_2
\end{Bmatrix}
\begin{Bmatrix}
F & s_1 & N\\
S' & i & s_2
\end{Bmatrix}.
\label{eq:hfpsi}
\end{align}
Through Eq.~\eqref{eq:hfpsi} we can conclude that states with different $N$ do not couple through the hyperfine interaction. The quantum number $N$ is preserved under the hyperfine Hamiltonian. However, the electronic total spin $S$ is not preserved. Singlet and triplet states do couple due to this interaction for they have a non-zero matrix element.

Finally we study the matrix elements of the Fermi pseudopotential. In this notation, we have for the $\bar{L}, \bar{S}$ Fermi pseudopotential matrix element
\begin{align}
\langle \Psi_{\xi} | \hat{V}_{\bar{L}\bar{S}} | \Psi_{\xi'} \rangle=& 2 \pi (2 \bar{L}+1) \frac{\boldsymbol{\Upsilon}^{L,M_L}_{\bar{L} M_L}(n,R) \boldsymbol{\Upsilon}^{L',M_L'}_{\bar{L},M_L'}(n',R)}{\sqrt{  \boldsymbol{\Upsilon}^{L M_L}_{L M_L}(n,R)  \boldsymbol{\Upsilon}^{L' M_L'}_{L' M_L'}(n',R)}} \, \delta_{S \bar{S}} \, \delta_{S' \bar{S}} \nonumber \\
&\times \sum_{m_i} \sum_{J,M_J} C_{S M_J-M_L,i m_i}^{N \Omega-M_L} C_{S M_J-M_L',i m_i}^{N' \Omega-M_L'}   C_{\bar{L} M_L, S M_J-M_L}^{JM_J} C_{\bar{L} M_L', S M_J-M_L'}^{JM_J} a(\bar{L},\bar{S},J).
\label{eq:Fermipsi}
\end{align}
We can think of Eq.~\eqref{eq:Fermipsi} as a simple expression for the Fermi interaction matrix element that is proportional to the overlap matrix of the states with the corresponding $\bar{L}$ scattering channels multiplied by an effective dispersion length (volume). Also, from Eq.~\eqref{eq:Fermipsi} we immediately see that singlet and triplet states do not couple.  We can deduce that $s-$wave interaction does not couple trilobite and butterfly states, the only coupling between these two types of states arise from $p-$wave interaction. To study this coupling, we consider two cases for the total Fermi scattering interaction $\hat{V}_{\mathcal{F}}=\sum_{\bar{L},\bar{S}} \hat{V}_{\bar{L}\bar{S}}$. First, Potassium to exemplify the behavior when the $^3P_J$ splittings are negligible, and then Rubidium to understand the effects of these splittings. 

\subsection{Potassium}
\begin{figure}[ht!]
	\centering	
	\subfloat{%
		\includegraphics[width=0.4\columnwidth]{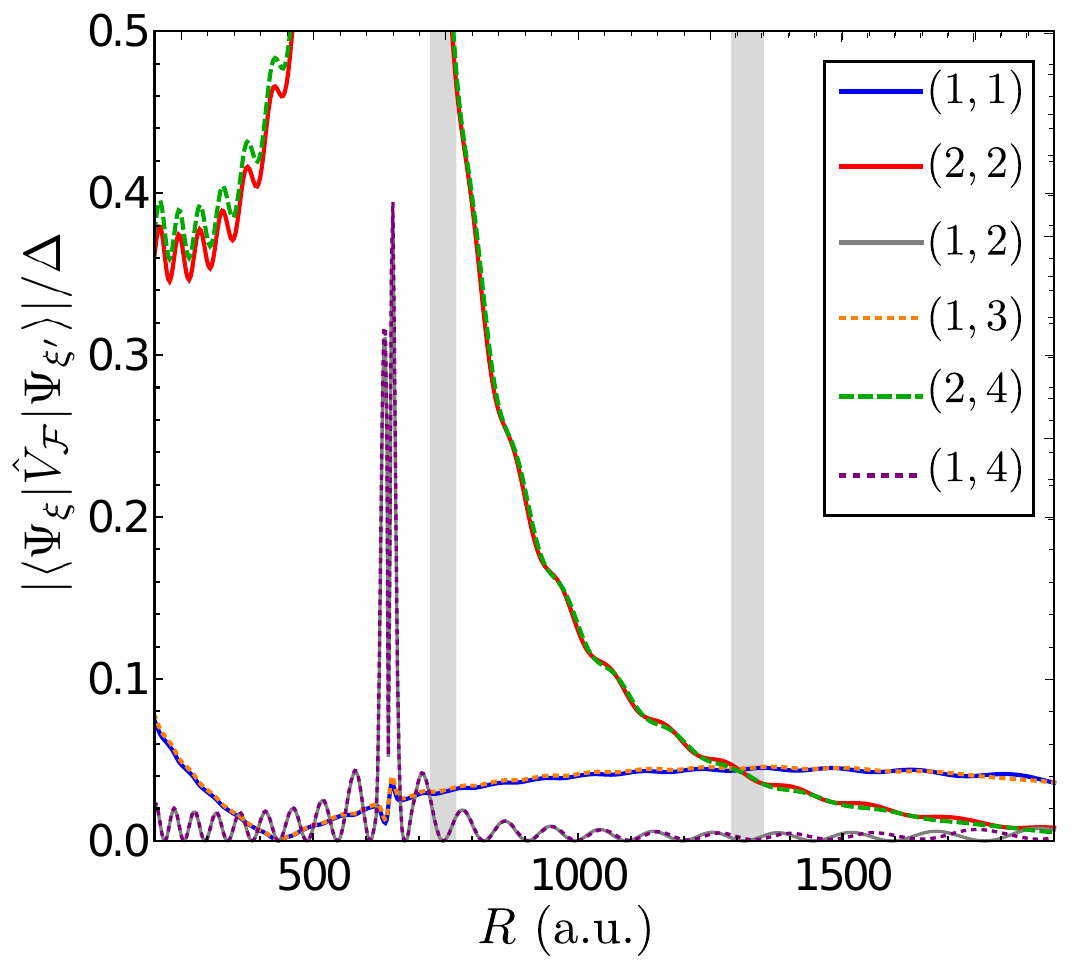}}
	\caption{Triplet Fermi pseudopotential matrix elements for $^{39}$K between $k$ and $k^\prime$ states $(k,k')$ as a function of the internuclear distance $R$ in $\Omega=3/2$ block. The basis states are distinguisheded by $\{n, L, M_L \}$; labels 1, 2 correspond to $\{34,0,0 \}$, $\{34,1,0\}$ and 3, 4 to $\{33,0,0 \}$, $\{33,1,0\}$  respectively. As established in the main text the results are independent of $N$ and we use $\Delta$ as normalization. The shaded region corresponds to the neighborhood of the avoided crossings.}	
	\label{fig:matrix_elementsK}
\end{figure}

When $a(\bar{L},S,J)$ does not depend on $J$, it can be taken out of the sum in Eq.~\eqref{eq:Fermipsi} and the sum of Clebsch-Gordan coefficients can be carried out explicitly. It results in $\delta_{N N'} \, \delta_{M_L M_L' }$, implying that the Fermi pseudopotential only couples trilobite and butterfly $|\Psi_{\xi} \rangle$ states with the same symmetry given by $N$ and $M_L$ and the matrix element is the same for all different $N$ values. As said before, for $^{39}$K there is a $^3P_J$ splitting that is negligible compared to other energies involved , and therefore $a(\bar{L},S,J)$ can  then be considered as approximately $J$-independent. In Fig. \ref{fig:matrix_elementsK} we present the non-zero matrix elements of the triplet Fermi pseudopotential for $n=34,35$ for this alkali atom. We focus on triplet states (interaction) as these are the dominant states in the corresponding bound LRRM. We see that the non-diagonal elements (associated to the coupling between trilobite and $M_L=0$ butterfly) are always much smaller than the diagonal matrix elements for internuclear distances that correspond to the region where bound trilobites molecules exist ($R \approx 900 \,a_0-2100\, a_0$) for both $n'=n$ and $n'=n\pm 1$. For this case we shall not observe considerable Fermi pseudopotential coupling between trilobite and butterfly states in accordance to the numerical diagonalization results for Potassium.

From Fig. \ref{fig:matrix_elementsK} we can also see that the Fermi matrix elements for $n'=n$ and $n'=n\pm 1$ are of comparable magnitudes. However, when considering the full Hamiltonian in a simple two level model we find that to first order, the $n\pm 1$ correction for the $n$ dominated state is given by $\langle \Psi_{\xi} | \hat{V}_{\mathcal{F}} | \Psi_{\xi'} \rangle/\Delta$ where $\Delta= |E^{\mathrm{Ryd}}_{n \ell\geqslant \ell_{\mathrm{min}}}- E^{\mathrm{Ryd}}_{n \pm 1 \ell\geqslant \ell_{\mathrm{min}}}|$. We see that for trilobite states this correction is always small $$\langle n=34,L=0,M_L=0\vert \hat{V}_{\mathcal{F}}\vert n=33,L=0,M_L=0\rangle/\Delta\leqslant 0.05$$ in the region of interest. On the other, for butterfly states in the bound region $R\approx 200 \,a_0-400 \,a_0$ this correction is more significant $$\langle n=34,L=1,M_L=0\vert \hat{V}_{\mathcal{F}}\vert n=33,L=1,M_L=0\rangle/\Delta \approx 0.4\, .$$ For this reason, it was already recognized in the numerical analysis that the contribution of $n \pm 1$ terms is not negligible for butterfly states.

\subsection{Rubidium}
\begin{figure}[ht!]
	\centering	
	\subfloat{%
		\includegraphics[width=0.4\columnwidth]{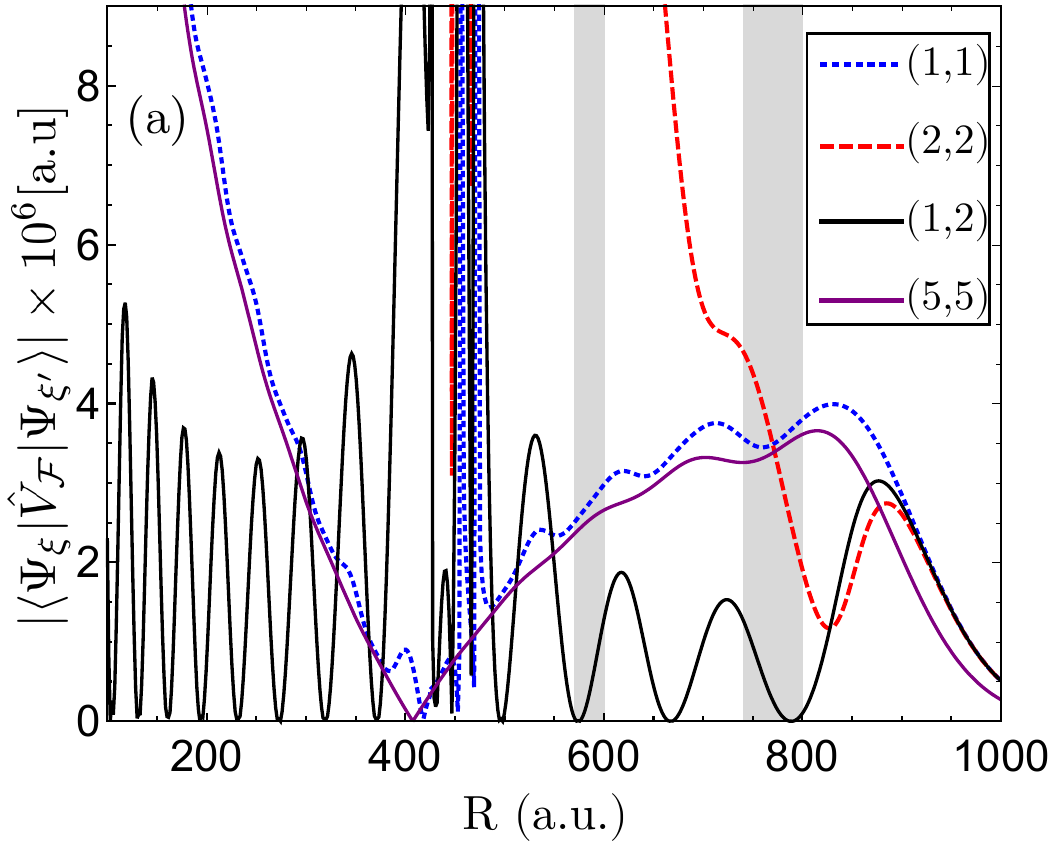}
		\label{fig:mat_eleA}
	} \hspace{0.1cm}
	\subfloat{%
		\includegraphics[width=0.4\columnwidth]{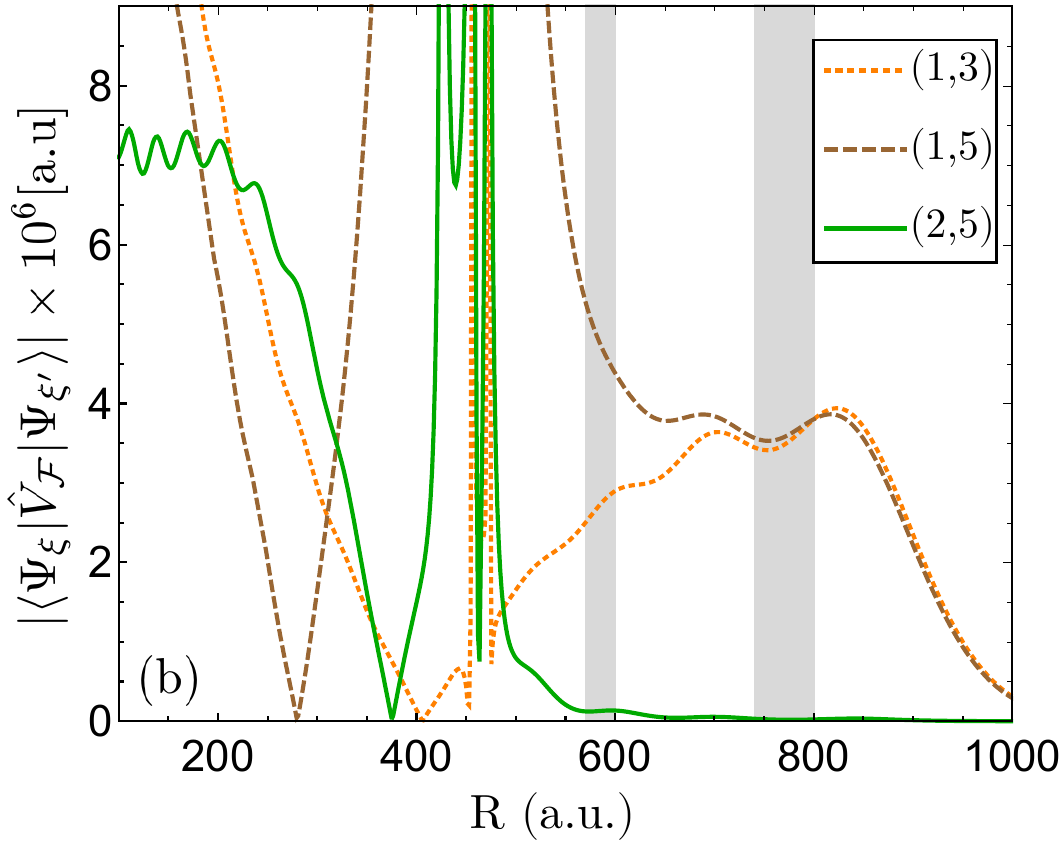}
		\label{fig:mat_eleB}
	}	
	\caption{Triplet Fermi pseudopotential matrix elements as a function of the internuclear distance $R$ between $k$ and $k^\prime$ states $(k,k')$ for (a) $N=N'$ and (b) $N \neq N'$ and the same principal quantum number $n=n'=22$ in $^{87}$Rb molecules. The basis states are distinguished  by $\{L,M_L,N\}$; labels 1,2, ...,5 corresponds to $\{0,0 ,{\scriptstyle \frac{3}{2}} \}$, $\{0,0 ,{\scriptstyle \frac{5}{2}} \}$, $\{1,0, {\scriptstyle \frac{3}{2}} \}$, $\{1,0, {\scriptstyle \frac{3}{2}} \}$, $\{0,0, {\scriptstyle \frac{5}{2}} \}$,$\{1,0, {\scriptstyle \frac{5}{2}} \}$ and $\{1,1, {\scriptstyle \frac{1}{2}} \}$ respectively. The shaded region corresponds to the neighborhood of the avoided crossings.}	
	\label{fig:matrix_elements}
\end{figure}

As we have exemplified in preceding Sections, the $N$-symmetry can be broken when the Fermi pseudopotential entails a strong magnetic interaction between the orbital angular momentum with respect to the ground state atom $L$ and the electronic spin $S$. This is the case for alkali atoms like Rb and Cs. For these atomic species, the $^3P_J$ splittings are considerable and cannot be ignored. In this case we find a non-zero matrix element between states with different $N$ and $M_L$. As example we consider $^{87}$Rb$_2$ molecules with $n=22$ and $\Omega=3/2$. The trilobite and butterfly avoided crossings are shown in the PECs of Fig. \ref{fig:NRb}. For this case, in Fig. \ref{fig:mat_eleA} we show some $N$ diagonal matrix elements for different $(L,M_L)$. For comparison we present matrix elements for states with $N'\neq N$ in Fig. \ref{fig:mat_eleB}. As we can see, these  off-diagonal matrix elements can be comparable or even larger than the diagonal elements, resulting in a significant coupling between different classes of states $\{L,M_L,N\}$ for the same principal quantum number $n$. 

\begin{figure}[ht!]
	\centering	
	\subfloat{%
		\includegraphics[width=0.36\columnwidth]{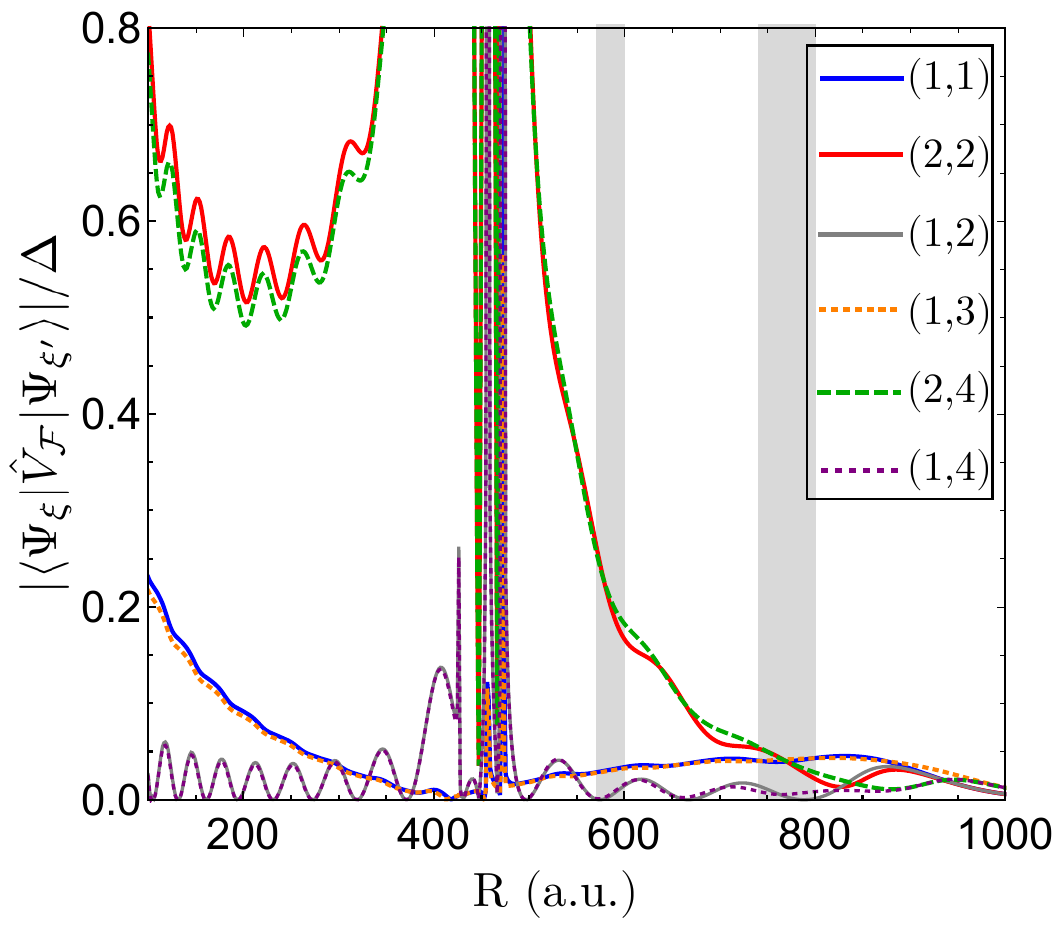}}
	\caption{Triplet Fermi pseudopotential matrix elements as a function of the internuclear distance $R$ for $^{87}$Rb between $k$ and $k^\prime$ states $(k,k')$ in $\Omega=N=3/2$ block. The basis $k$-states are distinguished by $\{n, L,M_L \}$ numbers; labels 1, 2 correspond to $\{22,0,0 \}$, $\{22,1,0\}$ and 3, 4 to $\{23,0,0 \}$, $\{23,1,0\}$  respectively. The shaded region corresponds to the neighborhood of the avoided crossings.}	
	\label{fig:matrix_elementsRb}
\end{figure}

To study the mixing of states with different principal quantum number we realize a similar analysis to the one for Potassium. In a set of the same $N=3/2$ states we obtain the Fermi pseudopotential matrix elements and compare them to $\Delta$ as shown in Fig.~ \ref{fig:matrix_elementsRb}. For trilobite states the correction is negligible, $$\langle n=22,L=0, M_L=0 \vert \hat{V}_{\mathcal{F}}\vert n=23,L=0, M_L=0\rangle/ \Delta \leqslant 0.1$$ in $R\approx 500 \,a_0-1000 \,a_0$. Meanwhile the correction $$0.5\leqslant \langle n=22,L=1, M_L=0 \vert \hat{V}_{\mathcal{F}}\vert n=23,L=1, M_L=0\rangle/ \Delta \leqslant 0.7$$ for butterfly states in $R\approx 100 \,a_0-400 \,a_0$ cannot be neglected. Again we find that  mixing between $n$ and $n\pm 1$ states only occurs for butterfly states. 

Despite the fact that matrix elements between high$-\ell$ $N$ and $N'$ can be different from zero, it must be noted that this quantum number has limited options. In our examples, $N=1/2, 3/2, 5/2$, so it is relatively simple to track couplings between different $N$ values by using the compact spin-orbital basis. In this  approach we  include both high$-\ell$ states for $n, n\pm 1$ and also low$-\ell$ states, particularly relevant for the $p-$wave interaction. For consistency, we also use low-$\ell$ states with well-defined $N$ and $S$. A similar analysis can be done for the coupling between high and low$-\ell$ states. We found that while the non-integer part of the $s$ state quantum defect is negligible, the corresponding $\Delta$ now defined as
$\Delta=E^{\mathrm{Ryd}}_{n \ell\geqslant \ell_{\mathrm{min}}}-E^{\mathrm{Ryd}}_{n', 0, 1/2}$ will always be very large compared to the Fermi pseudopotential matrix element. Resulting in no significant mixing of $n'$s states with $n$ trilobites. However, for $p-$wave states the Fermi matrix element can take higher values and quasi-degeneration between high and low-$\ell$ states is not required to observe the mixing of $n'p$ states with $n$ butterflies. This is consistent with the reported results here and in other studies \cite{Eiles2017,Eiles2018,Niederprum2016nature} for different atomic species.

In the same way as in the spin independent case \cite{Eiles2019}, the eigenstates of the Hamiltonian can be written as a linear combination of $|\Psi_{\xi}\rangle$ with a low$-\ell$ contribution. Since the spin trilobite states are not orthogonal, the problem results in a generalized eigenvalue equation that includes the full overlap matrix $\Lambda=\delta_{nn'}\delta_{N N'}\delta_{S S'} \boldsymbol{\Upsilon}^{LM_L}_{L'M_L'}(n,R)$.

\section{Spectroscopic consequences of the quasi symmetry generated by $\vec N$}
Our compact basis allows us to understand some spectroscopic properties of LRRMs. As the dipole moment operator acts only on the spatial degrees of freedom of the electron, it does not affect the spin component of the state. Therefore the $R$-dependent dipole moment matrix element between two elements of our compact basis is  
\begin{equation}
\vec{\mathcal{D}}(R)=\langle \Theta_{LM_L}^{(n)}|\hat{\vec{d}}|\Theta_{L'M_L'}^{(n')}\rangle \, \delta_{S S'} \, \delta_{N N'} \, \delta_{\Omega-M_L, \Omega'-M_L'}.
\label{eq:dipole_trilo}
\end{equation}
Expressing the dipole moment vector $\hat{\vec{d}}$ using the spherical base $\lbrace \hat{d}_0, \hat{d}_{\pm} \rbrace$ and from Eq.~\eqref{eq:dipole_trilo} we obtain a series of selection rules for given spin-orbitals $|\Psi_{\xi}\rangle$, Eq.~\eqref{eq:psiNcompleto}. First, the ones involving the electronic $S$ and molecular spin $N$ associated to the Kronecker deltas that appear in Eq. \eqref{eq:dipole_trilo}. These selection rules apply for all components of the dipole moment operator. Second, the selection rules that will be different for parallel or transverse (in the molecular frame) to the internuclear axis transitions. These selection rules arise as a combination of the projection condition $\Omega-M_L=\Omega'-M_L'$ and the spatial matrix element. For parallel transitions ($\hat{d}_0$), the spatial matrix elements requires $M_L=M_L'$ to be non-zero. This implies that $\Omega=\Omega'$. For perpendicular transitions ($\hat{d}_{\pm}$) the condition imposed by the spatial component is $\Delta M_L=\pm 1$ and according to Eq. \eqref{eq:dipole_trilo} this implies that $\Delta \Omega= \pm 1$.  

These selection rules can be used to understand the numerical results. When considering a bound trilobite LRRM in which the numerical state can be accurately approximated by only one triplet spin-orbital $|\Psi_{\xi}\rangle$ with $L=M_L=0$, only the $\hat{d}_0$ component is different from zero and only transitions in the same $\Omega$ block are allowed. This is precisely the result found with numerical calculations as the only trilobites in $(\Omega',n')$ with non-zero numerical transition dipole moments are  the ones with $n \neq n'$, $N=N'$ and $S=S'$ within a $\Omega'=\Omega$ block.

When considering butterfly states a distinction has to be made according to the extent of the $N$-symmetry. If $N$ remains a good quantum number, $^{39}$K for example, in the region where bound states exist, the numerical states are expressed as a linear combination of butterfly spin-orbitals of different principal quantum $n,n\pm 1$ and low angular momentum $n^*p$ states ($n^*$ determined by the quantum defect $\mu_1$) with the same well-defined $N$. In this case, the most relevant selection rule is $N'=N$ for each of the spin-orbitals $|\Psi_{\xi} \rangle$ required to represent the LRRM state. In one set of PECs correlated to $n^*p$ asymptotes on the same $\Omega$ block, the only allowed transitions are between radial butterflies. The existence of two radial butterfly LRRMs with the same $N$ occurs due to the splitting caused by the inclusion of low$-\ell$ states. However, despite having the same $N$ we note that due to the spatial shift between the PECs, numerical simulations show that the overlap of the nuclear wave functions corresponding to bound states will be minimal, reducing the transition probability. Within this $\Omega$ block but for a target set of $n^{*'}p$ asymptotically correlated states the selection rules admit various radial-radial and angular-angular transitions to the equivalent state. These transitions have considerable Franck-Condon factors. For different $\Omega$ blocks the selection rules permit perpendicular transitions between radial and angular butterflies in the same set of PECs correlated to the $n^{*'}p$ asymptotes.

If the $N$-symmetry is broken, $^{87}$Rb for example, we can expect that butterfly LRRMs will exhibit a qualitatively different spectroscopy. Contrary to the well-defined $N$ case, here for the same $\Omega$ and $n^{*'}p$ asymptote correlated block, transitions between all different states have a nonzero probability of occurring.
As illustrated in Fig.~\ref{fig:projRb}, each state can have a nonzero contribution from different values of $N$ in considerable spatial regions. Therefore, there will regions where the $N$ overlap between the two butterfly states of the same block under consideration can reach no negligible values and this can lead to non-zero transition electric dipole moments. Given that the different dipole matrix elements $\langle \Theta_{LM_L}^{(n)}|\hat{\vec{d}}|\Theta_{L'M_L'}^{(n')}\rangle$ for fixed $n,n'$ have comparable magnitudes, the sum of the $N$-overlaps between two states provides a rough estimate of which of these transitions within the same block are more probable. For the three butterfly PECs considered in Fig. \ref{fig:NRb} and using the projectors values of Fig. \ref{fig:projRb} to obtain the total $N$-overlap, we found that near the potential minima of the lowest radial dominated PEC, the overlap between that state and the angular dominated PEC is greater of  that between the two radial dominated PECs or the higher radial dominated PEC with the angular dominated PEC. Thus, the corresponding dipolar transition is more probable. 

\section{Conclusions}
We have shown that our spin-orbitals basis set results in simple expressions for the matrix elements of each term of the full electronic-ground state Hamiltonian for different atomic species of alkali atoms. These expressions permit a clear identification of which degrees of freedom are relevant for each of the interactions of the Hamiltonian and which of them will couple due to those interactions. Even when the non-diagonal matrix elements are significant the basis provides a good compact representation of them. 

The presented theoretical scheme, was applied to two paradigmatic examples with emphasis on the question of the effect of both the principal quantum number $n$ and the $N$ -symmetry of the spin-orbitals on the LRRMs spectroscopy. The calculations allow to understand the behavior and differences both in complexity and in the associated photon energies (and even polarization) for $^{39}$K and $^{87}$Rb LRRMs dipole transitions. They are mainly due to the broken $N$-symmetry derived from $^3P_J$ scattering splitting, whose relevance has already been recognized in the literature \cite{Greene2023}, though the role of this symmetry has not been previously remarked.

In the cases described in detail in this work, the $n$ values selected for $^{39}$K are considerably higher than the ones for $^{87}$Rb. As mentioned, part of the breaking of the $N$-symmetry occurs when the PECs are nearly degenerate and exhibit avoided crossings. For larger $n$ values, the intensity of the couplings present on the Hamiltonian (Fermi, hyperfine) change and can lead to a greater separation between the different PECs, breaking degeneracy and possibly contributing to the preservation of $N$ as a good quantum number. 

As its is well documented \cite{Eiles2017,Greene2023}, LRRMs PECs present a problem of formal convergence and dependence on the size of the base used to describe them. The efficiency of our compact basis instead of the Rydberg basis in numerical diagonalization on the convergence of the PECs deserves further studies. Besides such a basis could be ideal for studying LRRM in external electric and magnetic fields. Future studies could include generalizations of $\vec{N}$ as a symmetry operator to incorporate the scattering channel. For example $\vec{N}'=\vec{N}+\vec{L}$ has the advantage of reducing to $\vec{N}$ for trilobite states and may be of greater relevance in butterfly states. Finally the spin compact basis provides a direct way of identifying the relevant interaction terms in the neighborhood of avoided crossings, which are important in understanding non-adiabatic effects on LRRMs.

\acknowledgements{This work was partially supported by DGAPA-PAPIIT IN-104523. We thank RGJ for discussions and
reviewing the manuscript.}

\appendix
\section{Fermi pseudopotential diagonalization}
\label{section:Fermi}
In this Section we present details in the diagonalization of the Fermi pseudopotential within our perturbative model.

\subsection{$s-$wave scattering}
Here we focus on trilobite states. From Eq.~\eqref{eq:AA} for $\mathcal{A}$, we notice that the Fermi pseudopotential for $s$-wave ($L=M_L=0=0$) only couples states with $|m_j|=1/2$, and its matrix element will be zero for states that do not satisfy this condition. As a consequence, we only need to consider the elements of $W_{n_0}$ whose projection is $m_j=\pm 1/2$. 

Now we proceed with the detailed example of the construction and diagonalization of the matrix for the $s-$wave Fermi pseudopotential in $W_{n_0}$ for $\Omega=3/2$ in $^{39}$K. 

Only 3 triads of projections $(m_j;m_2,m_i)$ satisfy the condition on $m_j$. Therefore, the subspace in which we must diagonalize the pseudopotential is $\widetilde{W}_{n_0}= \lbrace  \lbrace |n_0 \ell j {\scriptstyle \frac{1}{2}} \rangle | {\scriptstyle \frac{1}{2}} {\scriptstyle \frac{1}{2}} \rangle \rbrace, \lbrace |n_0 \ell j {\scriptstyle \frac{1}{2}} \rangle | {\scriptstyle - \frac{1}{2}} {\scriptstyle \frac{3}{2}} \rangle \rbrace,$  $\lbrace  |n_0 \ell j {\scriptstyle - \frac{1}{2}} \rangle | {\scriptstyle \frac{1}{2}} {\scriptstyle \frac{3}{2}} \rangle \rbrace \rbrace$, where each $\lbrace |n_0 \ell j m_j \rangle | m_2 m_i \rangle \rbrace$ is a set of $2(n_0-3)$ states considering all possible values of $l$ and $j$. Ordered in this way, the subspace $\widetilde{W}_{n_0}$ can be thought of as composed of 3 different subsets $\lbrace |n_0 \ell j m_j \rangle | m_2 m_i \rangle \rbrace$ characterized by the projections $(m_j;m_2 , m_i)$. By evaluating the matrix elements of $\hat{V}_{\mathrm{Fermi}}$ on the states of $\widetilde{W}_{n_0}$, we obtain a block matrix, in which each block is associated with the pair $(m_j; m_2 , m_i)$, $(m_j' ; m'_2 , m' _i)$ and will be proportional to one of the matrices $\mathbb{M}_k$. \\

If we explicitly write separately the contribution of singlet ($S=0$) and triplet ($S=1$) 
\begin{equation}
\mathbb{V}_{\mathrm{Fermi}}^s=\mathbb{V}_{\mathrm{sing}}^s+\mathbb{V}_{\mathrm{trip}}^s,
\end{equation}
we found that in $\widetilde{W}_{n_0}$ the matrix for the $s-$wave Fermi pseudopotential  is
\begin{equation}
\mathbb{V}_{\mathrm{Fermi}}^s=2 \pi a(0,0,0,k(R)) \begin{pmatrix}
\mathbb{O} & \mathbb{O}               & \mathbb{O}                  \\
\mathbb{O}   & \frac{1}{2} \mathbb{M}_1 & -\frac{1}{2} \mathbb{M}_3    \\
\mathbb{O}   &-\frac{1}{2} \mathbb{M}_3^T  & \frac{1}{2} \mathbb{M}_2   
\end{pmatrix}
+2 \pi a(0,1,1,k(R)) \begin{pmatrix}
\mathbb{M}_1 & \mathbb{O}               & \mathbb{O}                 \\
\mathbb{O}   & \frac{1}{2} \mathbb{M}_1 & \frac{1}{2} \mathbb{M}_3   \\
\mathbb{O}   &\frac{1}{2} \mathbb{M}_3^T  & \frac{1}{2} \mathbb{M}_2  
\end{pmatrix}.
\end{equation} 

It is straightforward to find the eigenvectors and eigenvalues of the Fermi pseudopotential matrices. The singlet matrix is a rank 1 matrix which has a single non-zero eigenvalue given by
\begin{equation}
\lambda_{000}(R)= 2 \pi a(0,0,0,k(R)) \sigma_{00}^{n {\scriptstyle \frac{1}{2}}}(R),
\end{equation}
and eigenvector
\begin{equation}
|v_1\rangle =\frac{1}{\sqrt{2}}  \left( |\alpha_{{\scriptstyle \frac{1}{2}}, 0 \, 0}^{(n_0)}\rangle |{\scriptstyle - \frac{1}{2}} {\scriptstyle \frac{3}{2}} \rangle-|\alpha_{{\scriptstyle - \frac{1}{2}}, 0 \, 0}^{(n_0)}\rangle |{\scriptstyle \frac{1}{2}} {\scriptstyle \frac{3}{2}} \rangle \right).
\label{eq:trilo_psing}
\end{equation}

On the other hand, the matrix of the triplet  $\mathbb{V}_{\mathrm{trip}}^s$ has rank 2 and a non-zero eigenvalue with degeneracy 2. The eigenvalue and their respective linearly independent eigenvectors are
\begin{equation}
\lambda_{011}(R)= 2 \pi a(0,1,1,k(R))  \sigma_{00}^{n {\scriptstyle \frac{1}{2}}}(R),
\end{equation}
and
\begin{equation}
|v_2 \rangle=|\alpha_{{\scriptstyle \frac{1}{2}}, 0 \, 0}^{(n_0)} \rangle |{\scriptstyle \frac{1}{2}} {\scriptstyle \frac{1}{2}} \rangle, \hspace{1.5cm}
|v_3\rangle = \frac{1}{\sqrt{2}}  \left( |\alpha_{{\scriptstyle \frac{1}{2}}, 0 \, 0}^{(n_0)}\rangle |{\scriptstyle - \frac{1}{2}} {\scriptstyle \frac{3}{2}} \rangle+|\alpha_{{\scriptstyle - \frac{1}{2}}, 0 \, 0}^{(n_0)}\rangle |{\scriptstyle \frac{1}{2}} {\scriptstyle \frac{3}{2}} \rangle \right).
\label{eq:vecfermi}
\end{equation}

If we consider $ Q_{LM_L}^{n \ell  j}(R)$ independent of $j$ the eigenvector of the singlet matrix is in the kernel of the triplet matrix, and viceversa. So we have
\begin{equation}
\mathbb{V}_{\mathrm{trip}}^s |v_1 \rangle=0, \hspace{2cm} \mathbb{V}_{\mathrm{sing}}^s |v_i\rangle=0, \ \ i=2,3 .
\label{eq:kernel}
\end{equation}
Since we are considering only high$-\ell$ states this is a reasonable assumption over $ Q_{LM_L}^{n \ell  j}(R)$. 
As a consequence of Eq.~\eqref{eq:kernel}, the contributions of the singlet and triplet can be viewed as independent terms; the set formed by the union of the independent eigenvectors $|v_i\rangle$ is identical to the set of eigenvectors of the total $s-$wave Fermi interaction. Their eigenvalues are also the same as the union of the eigenvalues of $\mathbb{V}_{\mathrm{sing}}^s$ and $\mathbb{V}_{\mathrm{trip}}^s$.

The two triplet states span a degenerate subspace, whose energy correction $\lambda_{011}(R)$ produces the characteristic trilobite PECs. 

The electronic states $| \alpha_{m_j,L M_L}^{(n)} \rangle$ are orthogonal with respect to the principal quantum number $n$ and projection $m_j$
\begin{equation}
\langle \alpha_{m_j,L M_L}^{(n)} | \alpha_{m_j' ,L' M_L' }^{(n')}\rangle=\delta_{n n' } \delta_{m_j m_j'}.
\end{equation}
However, states with the same principal quantum number and projection but associated with different scattering channels are not orthogonal to each other.

\begin{figure}[ht!]
	\centering
	\includegraphics[width=0.4\linewidth]{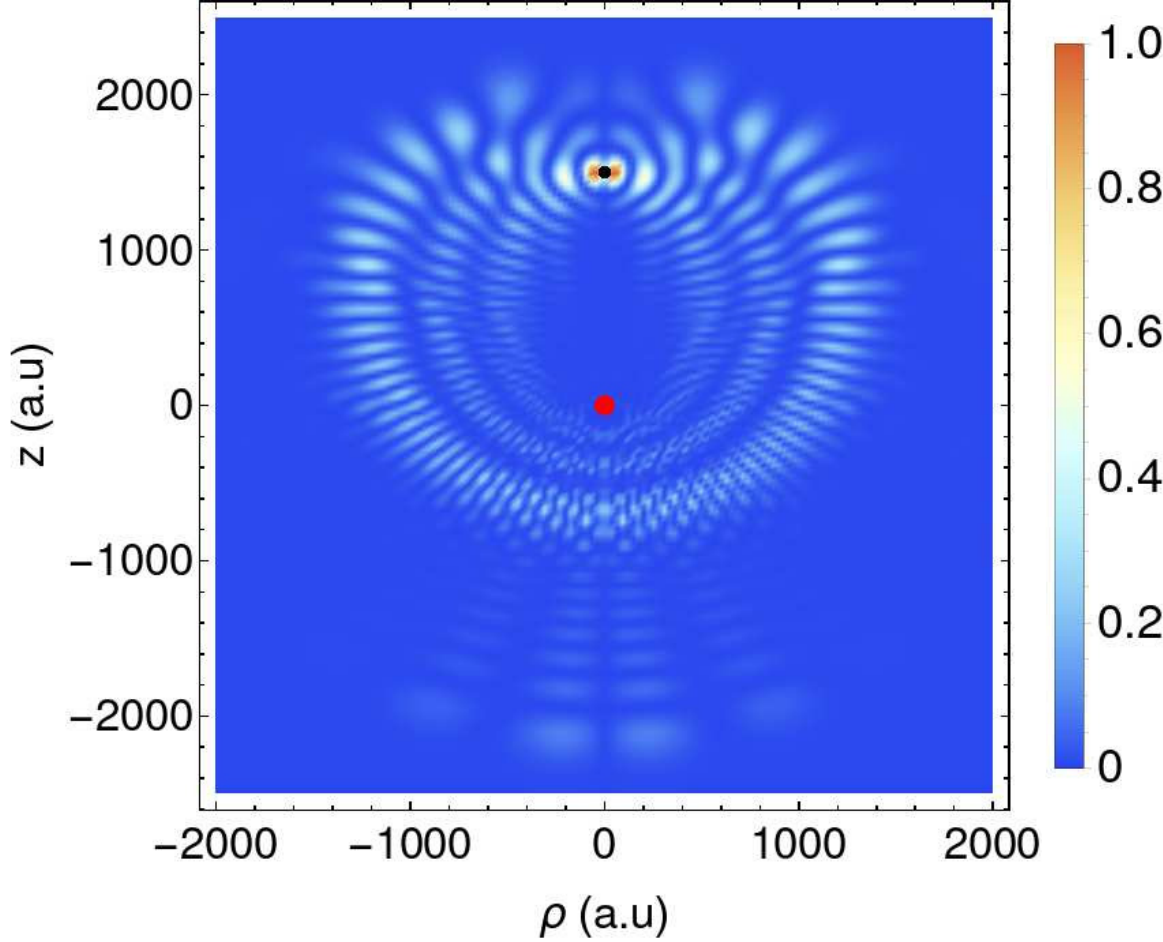}
	\caption{Trilobite unnormalized probability density $|\alpha_{{\scriptstyle \frac{1}{2}}, 0 \, 0}^{(34)}\rangle$ for $R=1495 \, a_0$ in cylindrical coordinates $(\rho,z)$. The Rydberg core and perturber atom are shown as red and black circles respectively.}
	\label{fig:tnatural}
\end{figure}

We will refer to the high angular momentum electronic states $| \alpha_{\pm {\scriptstyle \frac{1}{2}}, 0 \, 0}^{(n)} \rangle$ as trilobite \textit{fundamental} states. Fig.~\ref{fig:tnatural} shows the probability density for one of these states and the characteristic structure in the Rydberg electron wave function is evident. They depend parametrically on the internuclear distance $R$ through the functions $\widetilde{Q}$.

\subsection{$p-$wave scattering}
We now consider the $p-$wave scattering interaction for butterfly states with the same assumptions as in the case of $s$-wave scattering. We study the same $\Omega = 3/2$ case as for trilobite states. As shown in Fig.~\ref{fig:Kmariposaspin}, for this value of $\Omega$ there are at least eight potential curves with structure that we expect to correspond to butterfly PECs, apparently separated into two different classes of curves. The first class have a structure of several wells that could support vibrational levels. On the other hand, the second class are those PECs corresponding to a single well extended throughout the entire spatial region of interest.These eight curves arise almost exclusively from the Fermi interaction associated with the triplet-spin. With this in mind, we restrict our analysis to this case.

For $\Omega = 3/2$, we have 5 possible combinations of projections. As a consequence $W_{n_0}$ is composed of five subsets characterized by different projections and the Fermi pseudopotential matrix in this subspace $\mathbb{V}_{\mathrm{Fermi}}^{(p,1)}$ is a $5 \times 5$ block matrix.  We can write the matrix of the Fermi pseudopotential for $p-$wave triplet as a sum of contributions from different $J$. That is, we can write
\begin{equation}
\mathbb{V}_{\mathrm{Fermi}}^{(p,1)}=\mathbb{V}^{(p)}_{J=0}+\mathbb{V}^{(p)}_{J=1}+\mathbb{V}^{(p)}_{J=2}.
\label{eq:PJ}
\end{equation}

\begin{figure}[ht!]	
	\centering
		\includegraphics[width=0.8\columnwidth]{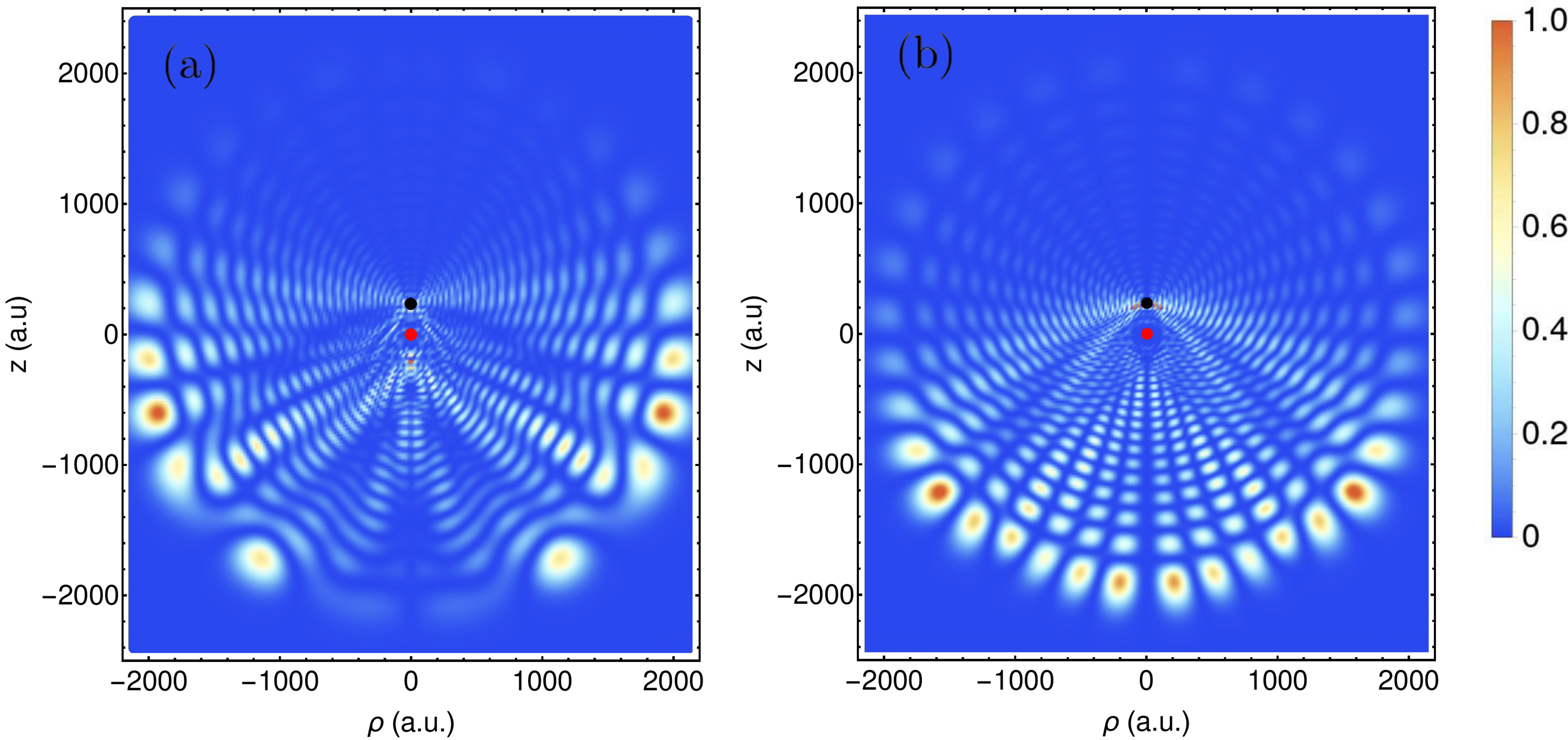}
	\caption{Butterfly unnormalized probability density for (a) radial $|\alpha_{{\scriptstyle \frac{1}{2}}, \,{1 \,0}}^{(34)} \rangle$ and angular $|\alpha_{{\scriptstyle \frac{1}{2}}, \,{1 \,1}}^{(34)} \rangle$ states for $R=230 \, a_0$ in cylindrical coordinates $(\rho,z)$. The Rydberg core and perturber atom are shown as red and black circles respectively.}	
	\label{fig:tipos_but}
\end{figure}

The expression given by Eq.~\eqref{eq:PJ} is valid in general. To proceed further, we make the additional assumption that the $^3P_J$ splitting of the phase shifts is negligible and all dispersion volumes $a(1,1,J)$ can be considered equals. We have seen that for Potassium this assumption is reasonable. By making this approximation, the sum over the possible values of $J$ is simplified and the terms that mix different $M_L$ cancel each other. It turns out that for this case we can separate the matrices according to their contributions of $M_L$. This is an important simplification as the matrices associated with each $M_L$ are independent in the same sense discussed in the  trilobite Section. For $p-$wave scattering, the Fermi pseudopotential does not completely break the degeneracy in $W_{n_0}$. Since it was possible to separate the contributions of $M_L=0$ and $|M_L|=1$, we can identify the molecules by approximate symmetry $\Sigma$ or $\Pi$. The eigenvectors of the matrix for $M_L=0$ are identified as radial butterfly states, and the ones for $|M_L|=1$ as  angular butterfly states. It should be emphasized that in the general case with different scattering volumes, it will not be possible to separate contributions from different $M_L$. Fig.~\ref{fig:tipos_but} shows the electronic probability density corresponding to the states associated with the two types of butterfly states.

For our study case $\Omega=3/2$ the matrix for $M_L=1$ has rank two while the matrix for $M_L=0$ has rank one. The resulting matrices have the same simple block structure as in the case of $s-$ wave scattering. Therefore finding their eigenvectors is straightforward. The radial and angular butterfly eigenstates are given by
	\begin{equation}
	|v_4\rangle = \frac{1}{\sqrt{2}}  \left( |\alpha_{{\scriptstyle \frac{1}{2}}, 0 \, 0}^{(n_0)}\rangle |{\scriptstyle - \frac{1}{2}} {\scriptstyle \frac{3}{2}} \rangle+|\alpha_{{\scriptstyle - \frac{1}{2}}, 1 \, 0}^{(n_0)}\rangle |{\scriptstyle \frac{1}{2}} {\scriptstyle \frac{3}{2}} \rangle \right), \hspace{1.5cm}
	|v_5 \rangle=|\alpha_{{\scriptstyle \frac{1}{2}}, 1 \, 0}^{(n_0)} \rangle |{\scriptstyle \frac{1}{2}} {\scriptstyle \frac{1}{2}} \rangle, 
	\label{eq:vecfermiP}
	\end{equation}
	and
	\begin{align}
		|v_6\rangle = &\frac{1}{\sqrt{2}}  \left( |\alpha_{{\scriptstyle \frac{3}{2}}, 1 \, 1}^{(n_0)}\rangle |{\scriptstyle - \frac{1}{2}} {\scriptstyle \frac{1}{2}} \rangle+|\alpha_{{\scriptstyle  \frac{1}{2}}, 1 \, 1}^{(n_0)}\rangle |{\scriptstyle \frac{1}{2}} {\scriptstyle \frac{1}{2}} \rangle \right), \\
		|v_7 \rangle=|\alpha_{{\scriptstyle  \frac{3}{2}}, 1 \, 1}^{(n_0)} \rangle |{\scriptstyle \frac{1}{2}} {\scriptstyle -\frac{1}{2}} \rangle, \hspace{1cm}
		&|v_8 \rangle=|\alpha_{{\scriptstyle - \frac{1}{2}}, 1 \, -1}^{(n_0)} \rangle |{\scriptstyle \frac{1}{2}} {\scriptstyle \frac{3}{2}} \rangle,  \hspace{1cm}
		|v_9 \rangle=|\alpha_{{\scriptstyle  \frac{1}{2}}, 1 \, 1}^{(n_0)} \rangle |{\scriptstyle -\frac{1}{2}} {\scriptstyle \frac{3}{2}} \rangle, 
		\label{eq:vecfermiP2}
	\end{align}
respectively.

\section{Hyperfine diagonalization}
\label{section:hyperfine}
Once the Fermi pseudopotential is diagonalized we still need to include the hyperfine term to approximately solve the Schroedinger equation of the full Hamiltonian written in Eq. \eqref{eq:hspin}. Taking advantage of the results from the previous Section, the general procedure is illustrated here for $\Omega=3/2$.
	
\subsection{Trilobite states} 
Since the singlet state is non-degenerate, its correction to the energy is simply the expected value of the hyperfine interaction. However the bound trilobite states come mostly from the triplet term, so we focus on this term.  The eigenstates of the triplet Fermi pseudopotential are degenerate. We use perturbation theory of degenerate states within the adequate subspace. In the illustrative case of study  ($\Omega=3/2$) the this degenerate subspace is spanned by  $\lbrace |v_2\rangle, |v_3 \rangle\rbrace$, so the hyperfine interaction matrix is a $2\times 2$ matrix. 

By diagonalizing the hyperfine interaction in the trilobite subspace, the degeneracy is completely broken and two different trilobite states are found. In the hyperfine coupled basis these triplet trilobite states are
\begin{equation}\label{eq:T1}
|\Psi_{\mathrm{T}1}^{(n_0)} \rangle_{\frac{3}{2}} = \sqrt{\frac{7}{10}} \,  |\alpha_{{\scriptstyle \frac{1}{2}} , 0 \,0}^{(n_0)}\rangle \left( \frac{5}{2} \sqrt{\frac{1}{7}}   |1 \, 1 \rangle +\frac{1}{2} \sqrt{\frac{3}{7}}   |2 \, 1 \rangle\right) -\sqrt{\frac{3}{10}} \,  |\alpha_{{\scriptstyle - \frac{1}{2}} , 0 \,0}^{(n_0)}\rangle   |2 \, 2 \rangle ,
\end{equation}

\begin{equation}\label{eq:T2}
|\Psi_{\mathrm{T}2}^{(n_0)} \rangle_{\frac{3}{2}}= \frac{2}{\sqrt{5}}  \, |\alpha_{{\scriptstyle \frac{1}{2}} , 0 \,0}^{(n_0)}\rangle |2 \, 1 \rangle+\frac{1}{\sqrt{5}} \, |\alpha_{{\scriptstyle - \frac{1}{2}} , 0 \,0}^{(n_0)}\rangle |2 \, 2 \rangle.
\end{equation}
whit respective energy corrections
\begin{equation}
\gamma_1=-\frac{1}{2} A_{\mathrm{HF}}, \hspace{1cm} \gamma_2=\frac{3}{4} A_{\mathrm{HF}}.
\label{eq:trilo3}
\end{equation}

For each value of $n_0$, perturbation theory predicts two different triplet trilobite states with energies 
\begin{equation}
E_i(R)=\epsilon_{n_0}(R)+2 \pi a(0,1,1,k(R)) \sum_{\ell,j} \left| \widetilde{Q}_{0 \, 0}^{n_0 \ell j {\scriptstyle \frac{1}{2}}} (R)\right|^2+\gamma_i.
\end{equation}

This prediction is consistent with the result of numerical diagonalization shown in Fig.~\ref{fig:Ktrilospin}. In Eqs.~\eqref{eq:T1} and \eqref{eq:T2} we have written the trilobite state vectors in a way that shows explicitly the natural orbitals of each vector. This allows for a direct comparison with the numerically obtained orbitals of Eq. \eqref{eq:trilo_nat1}.  A similar analysis provides the explicit trilobite states for each $\Omega$.

\subsection{Butterfly states} 
For $\Omega=3/2$ the radial butterfly vectors  $\lbrace |v_4\rangle, |v_5 \rangle\rbrace$ span a 2-dimensional degenerate subspace. After diagonalizing the hyperfine interaction we find two radial butterflies orbitals:
\begin{equation}
|\Psi_{\mathrm{B}1}^{(n_0)} \rangle_{\frac{3}{2}}= \sqrt{\frac{7}{10}} \, |\alpha_{{\scriptstyle \frac{1}{2}}, 1 \, 0}^{(n_0)} \rangle \left(  \frac{5}{2} \sqrt{\frac{1}{7}} \,  |1 \, 1 \rangle+\frac{1}{2} \sqrt{\frac{3}{7}}  \, |2 \, 1 \rangle    \right)-\sqrt{\frac{3}{10}} \,|\alpha_{{\scriptstyle - \frac{1}{2}}, {1 \,0}}^{(n_0)} \rangle |2  \,2 \rangle, \label{eq:B1}
\end{equation}

\begin{equation}
|\Psi_{\mathrm{B}2}^{(n_0)} \rangle_{\frac{3}{2}} = \frac{2}{\sqrt{5}}  \,  |\alpha_{{\scriptstyle \frac{1}{2}}, 1 \, 0}^{(n_0)}\rangle |2 \, 1 \rangle+\frac{1}{\sqrt{5} } \, |\alpha_{{\scriptstyle - \frac{1}{2}}, {1 \,0}}^{(n_0)}\rangle |2 \, 2  \rangle.\label{eq:B2}
\end{equation}

For the angular butterfly states we also must solve the hyperfine interaction in the degenerate subspace spanned by the four eigenvectors associated with $|M_L|=1$. Through a similar calculation to the one presented in the trilobite Section, we find the different angular butterfly states:
\begin{equation}\label{eq:B3}
|\Psi_{\mathrm{B}3}^{(n_0)} \rangle_{\frac{3}{2}}=\sqrt{\frac{1}{3}} \, |\alpha_{{\scriptstyle \frac{3}{2}} , 1 \,1}^{(n_0)} \rangle |1 \, 0 \rangle- \sqrt{\frac{2}{3}} \,|\alpha_{{\scriptstyle \frac{1}{2}} , 1 \,1}^{(n_0)} \rangle |1  \,1 \rangle, 
\end{equation}

\begin{equation}\label{eq:B4}
|\Psi_{\mathrm{B}4}^{(n_0)} \rangle_{\frac{3}{2}} = \sqrt{\frac{17}{30}} \,  |\alpha_{{\scriptstyle \frac{3}{2}} , 1 \,1}^{(n_0)}\rangle \left( \frac{5}{\sqrt{34}}   |1 \, 0 \rangle +\frac{3}{\sqrt{34}}   |2 \, 0 \rangle\right) +\sqrt{\frac{13}{30}} \,  |\alpha_{{\scriptstyle \frac{1}{2}} , 1 \,1}^{(n_0)}\rangle \left( \frac{5}{2} \sqrt{\frac{1}{13}}   |1 \, 1 \rangle -\frac{3}{2} \sqrt{\frac{3}{13}}   |2 \, 1 \rangle\right) ,
\end{equation}

\begin{equation}\label{eq:B5}
|\Psi_{\mathrm{B}5}^{(n_0)} \rangle_{\frac{3}{2}}= \sqrt{\frac{3}{5}}  \, |\alpha_{{\scriptstyle \frac{3}{2}} , 1 \,1}^{(n_0)}\rangle |2 \, 0 \rangle+\sqrt{\frac{2}{5}} \, |\alpha_{{\scriptstyle \frac{1}{2}} , 1 \,1}^{(n_0)}\rangle |2 \, 1 \rangle,
\end{equation}

\begin{equation}\label{eq:B6}
|\Psi_{\mathrm{B}6}^{(n_0)} \rangle_{\frac{3}{2}}= |\alpha_{{\scriptstyle - \frac{1}{2}} , 1 \,-1}^{(n_0)}\rangle |2 \, 2 \rangle.
\end{equation}

We note that in the same way as for the trilobite states, the perturber components in each eigenstate are the same for every value of $n_0$. The energy of $k_r-$th radial and $k_a-$th angular butterflies states are respectively
\begin{equation}
E_{\mathrm{B}k_r}(R)=\epsilon_{n_0}(R)+6 \pi a(1,1,1,k(R)) \sum_{\ell,  j} \left| \widetilde{Q}_{1 \, 0}^{n_0 \ell j {\scriptstyle \frac{1}{2}}} (R)\right|^2+\gamma_{kr}, 
\end{equation}

\begin{equation}
E_{\mathrm{B}k_a}(R)=\epsilon_{n_0}(R)+6 \pi a(1,1,1,k(R)) \sum_{\ell,  j} \left| \widetilde{Q}_{1 \, 1}^{n_0 \ell j {\scriptstyle \frac{1}{2}}} (R)\right|^2+\gamma_{ka},
\end{equation}

where $\gamma_k$ is the respective correction resulting of diagonalization of the hyperfine interaction in the degenerate butterfly sub-space.

According to the perturbative analysis, there should be four PECs associated with angular butterfly states. Studying the numerical eigenvector of each of the numerical PECs shown in Fig.~\ref{fig:Kmariposaspin} reveals that the four middle energy curves are the ones associated with angular butterfly states. 
For the case of radial butterfly states, perturbation theory predicts only two PECs, while in the numerical diagonalization we observe four energy curves. By analyzing the corresponding numerical eigenvector of these PECs, we find that they are mostly composed of radial butterfly states $n_0=34$, although contributions from $(n+2) p_j$ states  are also present. The splitting of each perturbative radial PECs in two different curves is a consequence of the admixture of butterfly and $np$ states.

A direct comparison between perturbative $|\Psi_{\mathrm{B}{k}}^{(n)}\rangle$ and full numerical natural orbitals shows that the latter cannot be written as a perturbative state of a single principal principal quantum number. However, we can use the perturbative states as a small basis to represent the numerical eigenvectors. For the angular butterflies the states can be written as butterflies of different $n_0$. For the radial butterflies is also necessary to include $np$ states.  Explicitly we can write the numerical angular buttefly associated to the $k_a-$th PEC as

\begin{equation}
| \Psi_{\Pi, k_a}^{(n)} \rangle_{\Omega}=A_{k_a}(R)|\Psi_{\mathrm{B}{k_a}}^{(n)}\rangle_{\Omega} +B_{k_a}(R)|\Psi_{\mathrm{B}{k_a}}^{(n+1)} \rangle_{\Omega}+C_{k_a}(R)|\Psi_{\mathrm{B}{k_a}}^{(n+1)} \rangle_{\Omega}.
\label{eq:mixn}
\end{equation} 

And the radial butterfly for the $k_r-$th PEC as
\begin{align}
|\Psi_{\Sigma,k_r} \rangle_{\Omega}=& \sum_{j,M_F} \left[ G_{j,M_F}^{(1)}(R) |(n+2) p_{j} \, \Omega-M_F \rangle |1 \, M_F \rangle+G_{j,M_F}^{(2)}(R)|(n+2) p_{j} \, \Omega-M_F \rangle |2 \, M_F \rangle \right]\nonumber \\
&A_{k_r}(R)|\Psi_{\mathrm{B}{k_r}}^{(n)}\rangle_{\Omega} +B_{k_r}(R)|\Psi_{\mathrm{B}{k_r}}^{(n)}\rangle_{\Omega}+C_{k_r}(R)|\Psi_{\mathrm{B}{k_r}}^{(n)}\rangle_{\Omega}
\label{eq:mixn2}
\end{align}

We note the mixing of different values for $n_0$ in the eigenvectors. This is a consequence of the $p-$wave interaction. For this we cannot neglect interaction between different $n$ states. 

\section{$\vec{N}$ projectors}
\label{section:Nproj}
To study further the $N$-symmetry, the expectation value of $\hat{P}_{N}$ can be calculated. Here,
\begin{equation}
\hat{P}_{N}=\sum_{M_N,S} \hat{P}_{N,M_N;S},
\end{equation}
is the projector on the state with good quantum number $N$ regardless of projection and electronic spin character. This analysis is necessary to achieve a better understanding of the observed value in $\langle \hat{N}^2\rangle$ presented in  Figs. \ref{fig:NK} and \ref{fig:NRb}. 

\begin{figure}[h]
	\centering	
		\includegraphics[width=0.9\columnwidth]{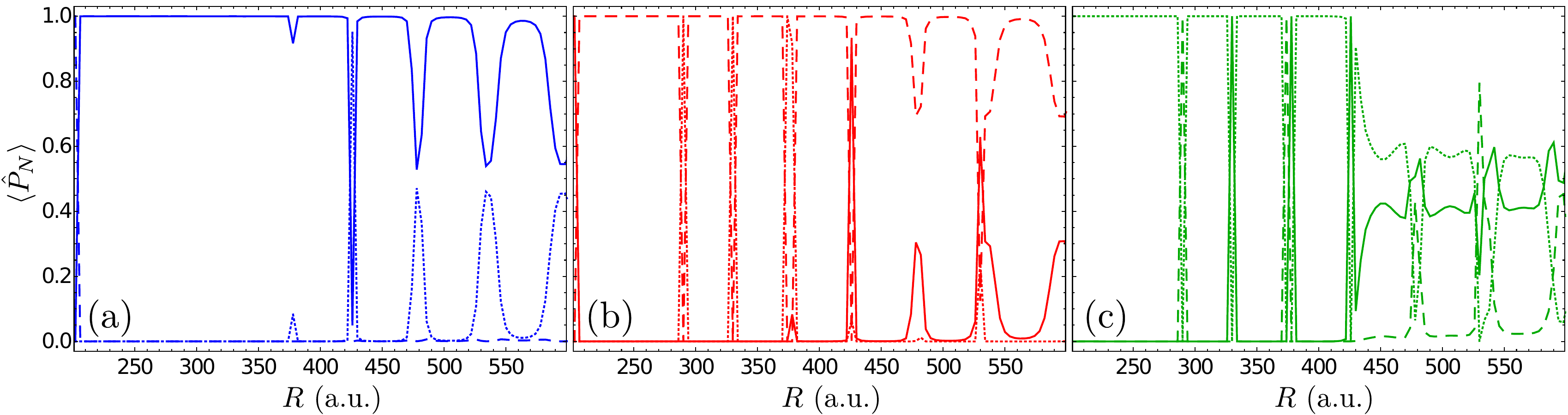}
	\caption{Expectation value of well-defined $N$ projectors $\hat{P}_{N}$ as a function of internuclear distance $R$ for butterfly states on $^{39}$K. (a), (b) and (c) correspond to the blue, red and green PECs of Fig. \ref{fig:NK}. $N=1/2,3/2,5/2$ are shown as dotted, continuous and dashed lines respectively.}	
	\label{fig:projK}
\end{figure}

\begin{figure}[h]
	\centering	
		\includegraphics[width=0.9\columnwidth]{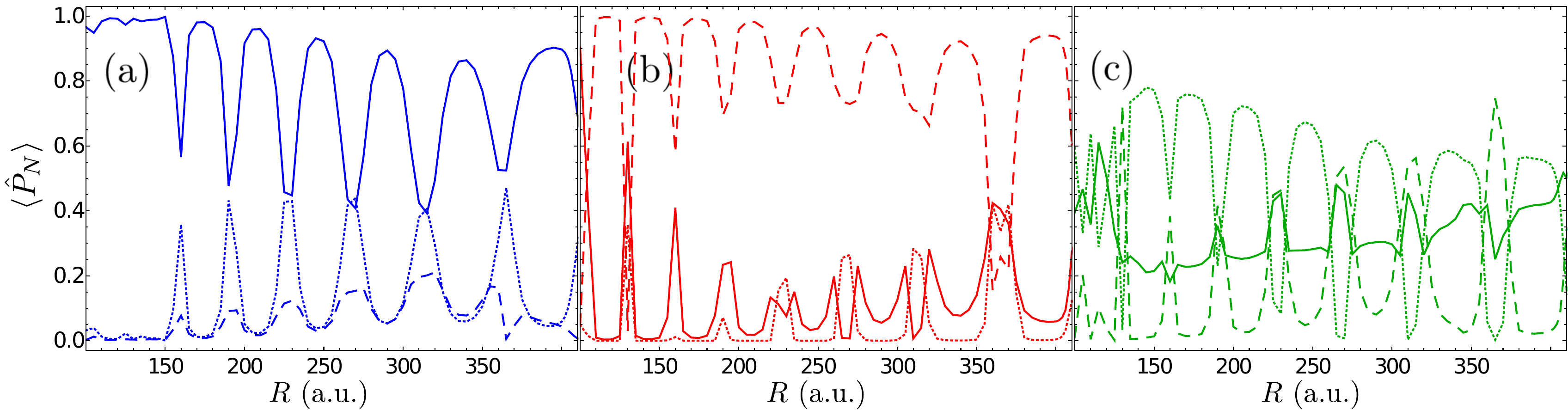}	
	\caption{Expectation value of well-defined $N$ projectors $\hat{P}_{N}$ as a function of internuclear distance $R$ for butterfly states on $^{87}$Rb. (a), (b) and (c) correspond to the blue, red and green PECs of Fig. \ref{fig:NRb}. $N=1/2,3/2,5/2$ are shown as dotted, continuous and dashed lines respectively.}	
	\label{fig:projRb}
\end{figure}

First, Fig. \ref{fig:projK} shows the projectors expectation values for $^{39}$K. We can see that except of a few singular points, corresponding to avoided crossings, for each of the butterfly PECs the expectation value of $\hat{P}_N$ for a given $N$ reaches the value of 1 and  remains practically constant around this value. On the contrary, for $^{87}$Rb contributions from all three possible values of $N$ are observed in Fig. \ref{fig:projRb} and $N$ no longer labels each PEC.

\bibliographystyle{unsrt} 
\bibliography{references}

\end{document}